\documentclass[11pt]{article}

\usepackage[T1]{fontenc}
\usepackage{lmodern}
\usepackage{microtype}
\usepackage{geometry}
\usepackage{amsmath,amssymb,amsthm}
\usepackage{array,booktabs,multirow}
\usepackage{algorithmic}
\usepackage[ruled,vlined,linesnumbered]{algorithm2e}
\usepackage{xurl}
\usepackage{tikz}
\usepackage[hypertexnames=false,colorlinks=true,citecolor=blue,linkcolor=blue,urlcolor=black]{hyperref}
\usetikzlibrary{positioning, arrows.meta}
\hypersetup{
  pdftitle={Improved Deterministic Algorithms for k-SAT: Generalization of Chains and Recurrence Proof},
  pdfauthor={group},
  pdfsubject={An alternate recurrence proof for Liu's deterministic 3-SAT chain analysis}
}

\allowdisplaybreaks

\title{Improved Deterministic Algorithms for $k$-SAT:\\ Generalization of Chains and Recurrence Proof}
\author{group}
\date{}

\newtheorem{theorem}{Theorem}[section]
\newtheorem{lemma}[theorem]{Lemma}
\newtheorem{corollary}[theorem]{Corollary}

\theoremstyle{definition}
\newtheorem{definition}[theorem]{Definition}

\theoremstyle{remark}

\newcommand{\br}{\textsf{BR}}
\newcommand{\dls}{\textsf{DLS}}

\newcommand{\ex}{\textsf{Expansion}}
\newcommand{\set}[1]{\{#1\}}
\newcommand{\pa}{\textsf{PA}}

\begin{document}

\author{
Binbin Cui\thanks{Shanghai Jiao Tong University, greenturtle@sjtu.edu.cn.}
\and
S. Cliff Liu\thanks{Shanghai Jiao Tong University, cliffliu@sjtu.edu.cn.}
\and
Yongrui Yu\thanks{Shanghai Jiao Tong University, ambiguity.yu@sjtu.edu.cn.}
}

\maketitle

\begin{abstract}

We present a deterministic algorithm for 3-SAT running in $1.32736^n$ time. The algorithm is built on a branching-and-chain framework augmented with a polynomial-time expansion procedure that joins short chains through additional clauses to form bounded-size blocks, which impose stronger constraints on the assignment space. After branching, the algorithm chooses between derandomized local search guided by the block structure and enumeration of block assignments together with selected port variables, reducing the remaining formula to 2-SAT. A weighted analysis of branching, partial-assignment, and local-search costs, together with inductive bounds for chains and block interactions, establishes the running-time guarantee. The framework is self-contained and also offers guidance for practical SAT solver design by identifying useful clause-overlap structures and suggesting when to switch between local-search and reduction-based strategies.
\end{abstract}

\section{Introduction}

$k$-SAT is one of the most fundamental NP-complete problems. Given a formula in Conjunctive Normal Form (CNF) with each clause having at most $k$ literals, the task is to find an assignment of Boolean values $1$(true) or $0$(false) to all variables, under which every clause contains at least one literal assigned to $1$. As a canonical NP-complete problem, $k$-SAT has driven decades of research on exponential-time algorithms.

For deterministic algorithms, the current best bound is due to OpenAI, who gives a sub-exponential bound but has not been fully verified as of the completion of this work. Prior to that work, the best upper bound was due to Liu, who gave an algorithm with running time strictly lower than $(2-2/k)^n$ for every $k\ge3$, and in particular $3^{n\log{\frac{4}{3}}/\log{\frac{64}{21}}}\approx1.32793^n$ for $k=3$ \cite{liu2018ksat}. This bound has stood for several years. At a high level, Liu's algorithm first attempt to reduce $F$ to a $(k-1)$-CNF within allowed time, and call the solver of $(k-1)$-SAT. Especially for $3$-CNF, it applies a branching algorithm to do this better. If it fails to do this within allowed time, the output of $\br$ can be used to narrow the search space of $\dls$.

\subsection{Brief Review of k-SAT}

The deterministic local-search algorithms most closely related
to our work developed along two directions: searching
efficiently near a given assignment, and using clause
structure to choose better starting assignments.

Sch\"oning~\cite{schoning1999probabilistic} introduced a
randomized local-search algorithm for $k$-SAT in 1999.
Starting from a uniformly random assignment, the algorithm
repeatedly chooses an unsatisfied clause and flips a uniformly
chosen variable in that clause. At least one such variable
differs from any fixed satisfying assignment, giving each
step a chance to move toward a solution. After $O(n)$ steps,
the algorithm restarts if no solution has been found.
On satisfiable formulas, the expected time to find a solution is
$\left(2-\frac{2}{k}\right)^n$.
For $3$-SAT, this gives the exponential base $4/3$, or the
rounded upper bound $1.33334$.

Hofmeister, Sch\"oning, Schuler, and Watanabe
(HSSW)~\cite{hofmeister2002probabilistic} improved this approach
by using disjoint clauses to guide the search.
The variant underlying the later derandomization constructs
a maximal family of pairwise variable-disjoint $3$-clauses.
If the family is small, it enumerates all assignments
satisfying these clauses. Maximality ensures that every
remaining $3$-clause contains an assigned variable, so each
residual formula is a $2$-CNF. If the family is large, the
algorithm instead uses a nonuniform distribution on its
satisfying assignments to initialize Sch\"oning's random
walk. These clause constraints improve the walk's success
probability. Balancing the two cases gives the
$1.3303^n$ bound used in the later derandomization.

Dantsin et al.~\cite{dantsin2002deterministic} developed a
deterministic approach based on covering codes.
They cover the assignment space by Hamming balls and search
each ball deterministically. This gives a general exponential
base of $2-2/(k+1)$, with a refined analysis yielding
$1.481$ for $3$-SAT. Moser and Scheder~\cite{moser2011full}
subsequently improved the deterministic search inside a
Hamming ball. Combined with covering codes for the starting
assignments, their method achieves running time $\left(\frac{2(k-1)}{k}\right)^{n+o(n)}$,
matching Sch\"oning's exponential base.

Makino, Tamaki, and Yamamoto
~\cite{DBLP:journals/algorithmica/MakinoTY13}
then derandomized the HSSW approach.
Their construction uses covering codes with varying radii
to accommodate the nonuniform assignment space induced by
the disjoint clauses. They search from these codewords using
Moser and Scheder's deterministic local search.
Together with enumeration in the small-family case, this
yields a deterministic $3$-SAT algorithm with running time
$O(1.3303^n)$.

Liu~\cite{liu2018ksat} extended this framework from disjoint
clauses to chains, in which only consecutive clauses may
share variables. The branching procedure either solves the
formula within its allotted time or produces enough chains
to make derandomized local search efficient.
Liu constructs generalized covering codes for the resulting
assignment space using distributions determined by linear
programming. This gives a deterministic $3$-SAT bound of
$O(1.32793^n)$ and improves the bounds for general $k$-SAT.
The combination of branching, clause structure, and
generalized covering codes provides the starting point
for our algorithm.

A separate line of randomized algorithms begins with
PPSZ~\cite{DBLP:journals/jacm/PaturiPSZ05}.
PPSZ processes variables in a random order, inferring their
values when possible and guessing otherwise.
Hertli~\cite{hertli20143} showed that the bounds previously
established for uniquely satisfiable formulas also hold for
general $3$-SAT and $4$-SAT.
More recently, Scheder~\cite{scheder2024ppsz} proved a strict
exponential improvement over the classical PPSZ analysis
for every fixed $k\ge3$, without changing the algorithm.
These results remain randomized.
Table~\ref{tab:ksat-comparison} includes them alongside the
deterministic bounds and identifies this distinction.

\subsection{Liu's Work on Deterministic 3-SAT}

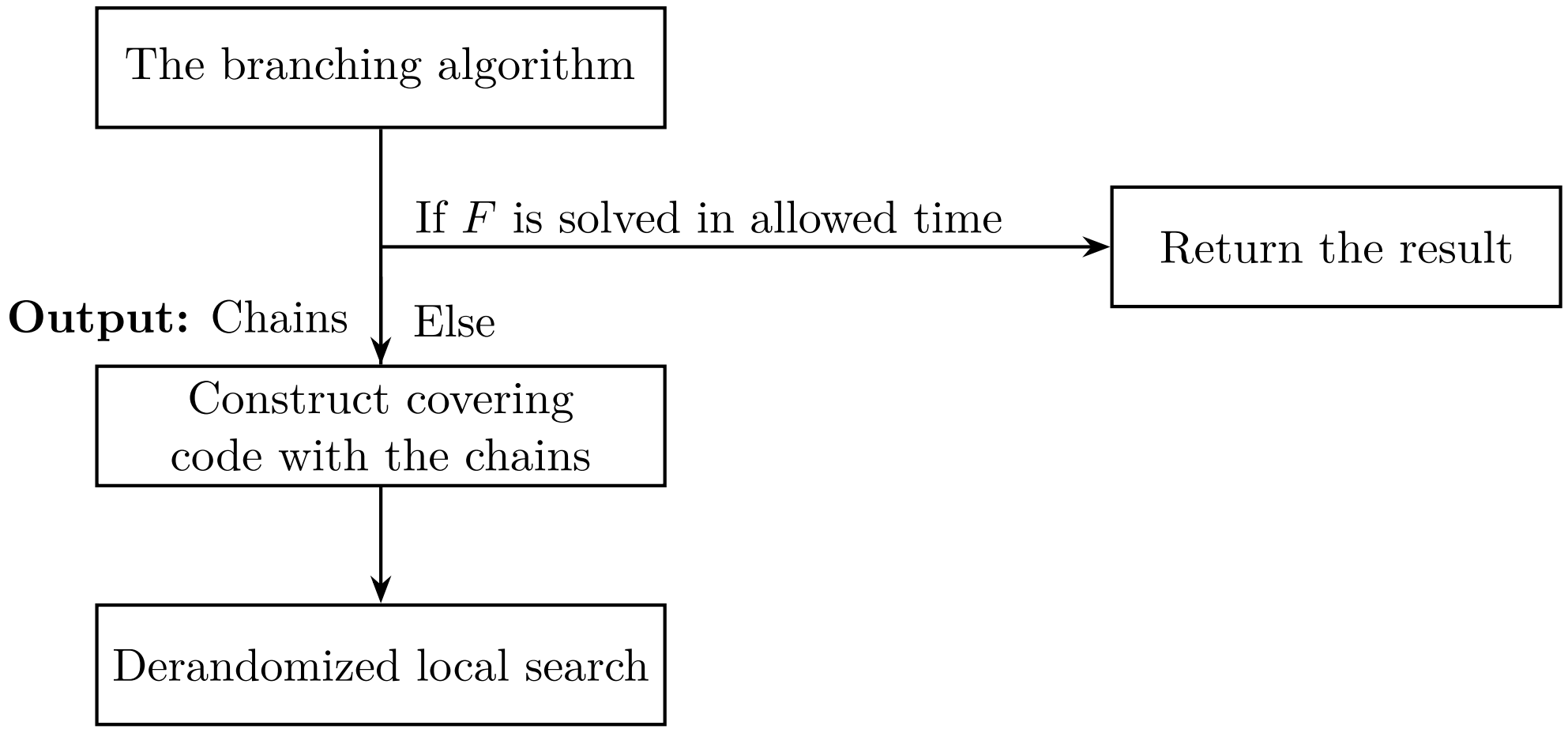
\begin{figure}[htbp]
    \centering
    \begin{tikzpicture}[
        block/.style = {rectangle, draw=black, thick, minimum height=1cm, text width=4.5cm, align=center, fill=white},
        arrow/.style = {->, >=Stealth, thick, draw=black}
    ]
        \node[block] (BR) at (0, 0) {The branching algorithm};
        \node[block] (CCC) at (0, -3) {Construct covering code with the chains};
        \node[block] (DLS) at (0, -5) {Derandomized local search};
        \node[block, text width=3.5cm] (Return) at (8, -1.5) {Return the result};
        
        \coordinate (Split) at (0, -1.5);
        
        \draw[thick] (BR.south) -- (CCC.north);
        \draw[arrow] (Split) -- node[pos=0.45, above, fill=white, inner sep=3pt] {If $F$ is solved in allowed time} (Return.west);
        
        \draw[arrow] (0, -1.75) -- (CCC.north) 
            node[pos=0.5, left=0.15cm, fill=white, inner sep=3pt] {\textbf{Output:} Chains}
            node[pos=0.5, right=0.15cm, fill=white, inner sep=3pt] {Else};
        
        \draw[arrow] (CCC.south) -- (DLS.north);
    \end{tikzpicture}
    \caption{The workflow of the algorithm of Liu}
    \label{fig:old_workflow}
\end{figure}

In 2018, Liu improved deterministic $3$-SAT by combining a branching algorithm with the derandomized local search. Its workflow is as Figure~\ref{fig:old_workflow}. This algorithm can be seen as a generalization of HSSW, and the main technique used is chain.

A chain is a clause sequence in which adjacent clauses have at least $1$ shared variable. Thus, unlike the disjoint clauses used by HSSW, a chain can capture insightful structure among overlapping clauses. Overlap relations between clauses in a chain restrict the search space of derandomized local search to a proper subset of the Cartesian product of the individual clause solution spaces. Because of the overlapping, the solution space of a chain is smaller than the product of the solution spaces of its individual clauses. Liu restricts the search space of derandomized local search in the product of the solution space of every constructed chain. This makes the search space smaller than that of HSSW. 

Chains are constructed by the branching algorithm. the branching algorithm constructs chains while attempts to reduce the formula to a $2$-CNF by a specific simplification method and a branching rule. If it succeeds to reduce within allowed branching steps, then the remaining formula can be solved in polynomial time. Otherwise, the algorithm constructs generalized covering code using the chains constructed by the branching algorithm.

Producing more chains reduces more search cost of derandomized local search but costs more branching time of the branching algorithm. Balancing the two gives a $3$-SAT running-time bound of $1.32793^n$. The improved $3$-SAT bound also feeds into Liu’s bounds for larger $k$.

\subsection{Our Contribution}

We observe that Liu's algorithm performs poorly when the branching algorithm fails to solve $F$ and produces too many short chains, which can only provide very loose constraints for the search space of derandomized local search. Therefore, a possible direction for improvement is to strengthen these constraints after running the branching algorithm.

Our method is to introduce an expansion procedure after the branching algorithm. Its task is to find more clauses in the original formula that can join the short chains together after running the branching algorithm. A greedy procedure will do this in polynomial time. Then, we reduce the formula to a $2$-CNF by enumerate partial assignments, or perform derandomized local search with the stronger constraints.

Analyzing the new algorithm is challenging. However, a simplified version of algorithm and analysis already yield a strict improvement on running time. We also have gained stronger improvement through a harder analysis. We believe this provides a promising route to further optimization, including more precise upper bound analysis and the introduction of more complex clause structures.

\paragraph{Practical Implication.} In practice, our framework provides a self-contained analysis of how clause-overlap structures affect the costs of local search and partial-assignment reductions, and thereby suggests directions for practical solver design. Studies of community structure show that interactions among clauses can help explain solver behavior and identify useful learned
clauses \cite{ansotegui2012community,ansotegui2015relevant}.
The chains and blocks studied here offer concrete structural features whose benefits can be quantified through their effects on enumeration and local-search costs. These features could also inform strategy selection: SATzilla demonstrates the value of choosing solvers according to instance features \cite{xu2008satzilla}, while hybrid solvers show
that conditions for invoking local search or systematic reasoning can affect practical performance \cite{audemard2009integrating,cai2021cooperation}.
Our comparison of the subroutine costs provides a theoretical basis for analogous choices between local search and partial-assignment reductions. Whether these criteria yield practical improvements remains an empirical question.

\paragraph{Relation to the recent claimed ETH refutation.}
Since the submission of this work, OpenAI has posted a manuscript claiming a deterministic $2^{o(n)}$-time algorithm for exact counting on fixed-$k$-CNF formulas, which would refute ETH if fully verified. Our result is logically independent of that claim. We give an explicit deterministic $1.32736^n$-time algorithm for $3$-SAT, improving Liu’s $1.32793^n$ bound through a different derandomization framework. The claimed subexponential result is still undergoing independent scrutiny, and its currently stated explicit $3$-SAT bound and hidden constants are not directly comparable with our explicit-base guarantee. Our contribution is therefore best understood as an explicit deterministic improvement and as a set of techniques for derandomizing local search algorithms.

\subsection{Our Concrete Results}

We view the expansion procedure and the analysis of clause structures as new conceptual ideas that can lead to larger improvements. This improvement is not limited to $3$-SAT; it also yields an improvement for $k\ge4$. Although the above result is far from optimal and the resulting improvement is numerically modest, it already yields a strict improvement, indicating that this direction is very promising. In this paper, we will show how $3$-SAT can be improved to run in $1.32736^n$ following our route.

\subsection{Our Main Technique}

\begin{figure}[htbp]
    \centering
    \begin{tikzpicture}[
        block/.style = {rectangle, draw=black, thick, minimum height=1cm, text width=4.5cm, align=center, fill=white, inner sep=4pt},
        arrow/.style = {->, >=Stealth, thick, draw=black}
    ]
        \node[block] (BR) at (0, 0) {The branching algorithm};
        \node[block] (Exp) at (0, -3) {Expand the chains to blocks}; 
        
        \node[block, text width=3.5cm] (Return) at (5.5, -1.5) {Return the result};
        
        \node[block] (CCC) at (-5, -5) {Construct covering code};
        \node[block] (RD) at (5, -5) {Reduce $F$ to a 2-CNF};
        \node[block] (DLS) at (-5, -7) {Derandomized local search};
        \node[block] (S2S) at (5, -7) {Solve $F$ in polynomial time};
        
        \coordinate (SplitBR) at (0, -1.5);
        \coordinate (SplitExp) at (0, -4);
        
        \draw[thick] (BR.south) -- (SplitBR);
        \draw[arrow] (SplitBR) -- (Return.west) node[pos=0.5, above, fill=white, inner sep=3pt] {If $F$ is solved};
        
        \draw[arrow] (SplitBR) -- (Exp.north) 
            node[pos=0.5, left=0.15cm, fill=white, inner sep=3pt, anchor=east] {\textbf{Output:} Chains}
            node[pos=0.5, right=0.15cm, fill=white, inner sep=3pt, anchor=west] {If $F$ is not solved};
        
        \draw[thick] (Exp.south) -- (SplitExp) 
            node[pos=0.6, left=0.15cm, fill=white, inner sep=3pt, anchor=east] {\textbf{Output:} Blocks};
        
        \draw[arrow] (SplitExp) -| (CCC.north);
        \draw[arrow] (SplitExp) -| (RD.north);
        
        \node[above, fill=white, inner sep=2pt] at (0, -4.65) {Choose the Cheaper One};
        
        \draw[arrow] (CCC.south) -- (DLS.north);
        \draw[arrow] (RD.south) -- (S2S.north);
    \end{tikzpicture}
    \caption{The workflow of our algorithm}
    \label{fig:improved_workflow}
\end{figure}
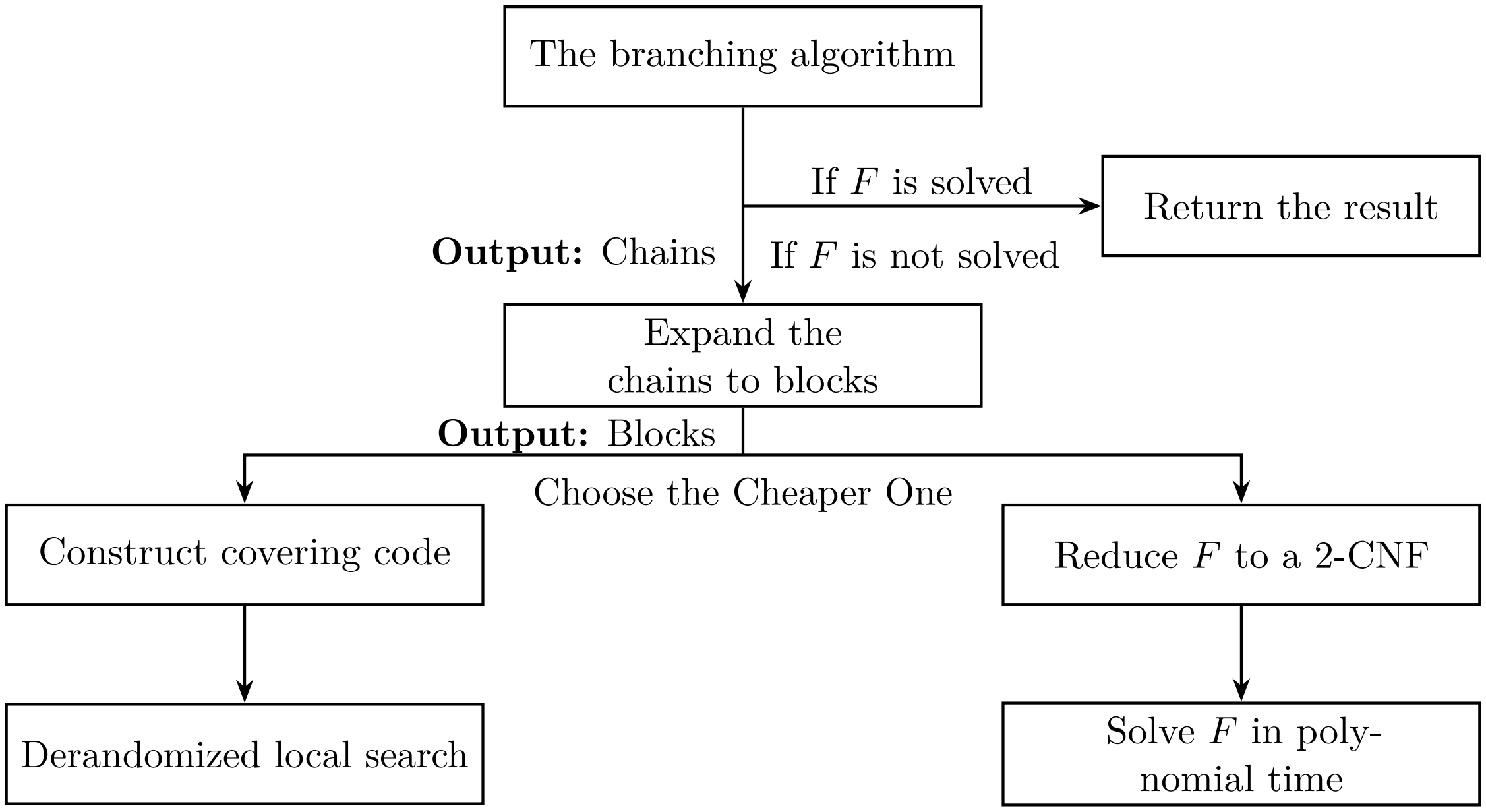

Our subsequent $1.32736^n$ algorithm builds on the base of the framework of Liu. The algorithm of Liu performs poorly if the branching algorithm fails in the reduction and constructs too many short chains. Specifically, we introduce a new algorithm after the branching algorithm, whose task is to improve this issue by expanding chains generated by the branching algorithm to more complex structures. This is the most significant difference between our algorithm and Liu's algorithm. In a word, our algorithm is as Figure~\ref{fig:improved_workflow}.

Chains are linear structures: each clause overlaps only with its neighbors in the chain. These overlaps allow a restriction imposed by one clause to influence the next, but a short chain has little room for such restrictions to accumulate. That is why short chains can only provide loose constraints. Intuitively, more intricate overlap relations between clauses should yield stronger constraints. If the individual short chains are connected into a larger structure by additional clauses, this structure will exhibit more complex overlap relations and thus provide stronger constraints. This led to blocks, which can be seen as generalized chains. 

A block can be viewed as a set of clauses, in which clauses are connected by overlap relations. Blocks can be seen as graphs if we see clauses as a vertices and overlap relations as edges. In a block, each of the $k$ variables in a $k$-clause can be used to form a overlap relation. By contrast, in a chain, each clause interacts only with its at most $2$ neighbors. Overlaps can link several chains and constrain their assignments jointly. Their more complex overlapping structure give them stronger ability to reduce the search space of derandomized local search.

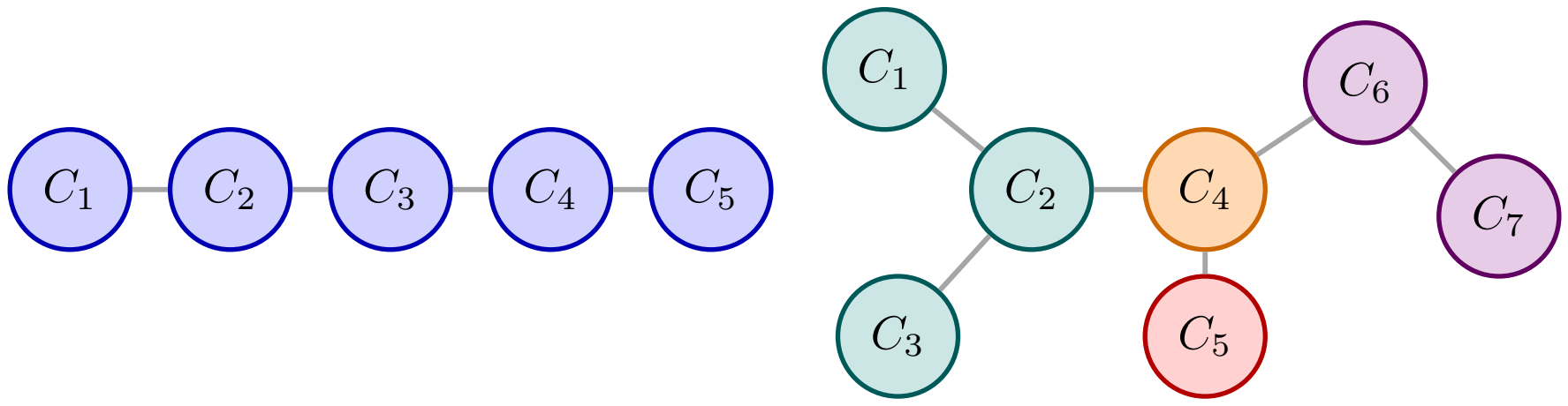
\begin{figure}[t]
\centering
\begin{tikzpicture}[
    clause/.style={
        circle, minimum size=9mm, inner sep=0pt,
        line width=1pt, font=\small
    },
    chainclause/.style={
        clause, draw=blue!70!black, fill=blue!18
    },
    blockchain1/.style={
        clause, draw=teal!70!black, fill=teal!20
    },
    blockchain2/.style={
        clause, draw=red!70!black, fill=red!18
    },
    centerclause/.style={
        clause, draw=orange!80!black, fill=orange!30
    },
    blockchain3/.style={
        clause, draw=violet!75!black, fill=violet!20
    },
    overlap/.style={draw=gray!70, line width=1.1pt}
]

\node[chainclause] (c1c) at (0,0) {$C_1$};
\node[chainclause] (c2c) at (1.2,0) {$C_2$};
\node[chainclause] (c3c) at (2.4,0) {$C_3$};
\node[chainclause] (c4c) at (3.6,0) {$C_4$};
\node[chainclause] (c5c) at (4.8,0) {$C_5$};
\draw[overlap] (c1c)--(c2c)--(c3c)--(c4c)--(c5c);
\node at (2.4,-2) {(a) Chain};

\node[blockchain1] (c1b) at (6.1, 0.9) {$C_1$};
\node[blockchain1] (c2b) at (7.2, 0.0) {$C_2$};
\node[blockchain1] (c3b) at (6.2,-1.1) {$C_3$};
\node[centerclause] (c4b) at (8.5, 0.0) {$C_4$};
\node[blockchain2] (c5b) at (8.5,-1.1) {$C_5$};
\node[blockchain3] (c6b) at (9.7, 0.8) {$C_6$};
\node[blockchain3] (c7b) at (10.7,-0.2) {$C_7$};

\draw[overlap]
    (c1b)--(c2b) (c2b)--(c3b) (c2b)--(c4b)
    (c4b)--(c5b) (c4b)--(c6b) (c6b)--(c7b);

\node at (8.1,-2) {(b) Block};
\end{tikzpicture}
\caption{Schematic clause-overlap structures for chains and blocks. Each vertex is a clause, and each edge indicates that its two clauses share at least one variable. The clauses in the same color denote a chain.}
\label{fig:chains-and-blocks}
\end{figure}

Blocks are constructed greedily by $\ex$. $\ex$ scans all chains constructed by the branching algorithm and examines how other clauses overlap them. When several clauses overlaps with several chains, $\ex$ can form a block containing the clauses and all chains touched by them. In this way, $\ex$ brings together short chains that would otherwise be considered separately, allowing the clauses connecting them to constrain their assignments jointly. $\ex$ runs in polynomial time, so it would not affect the upper bound on the time complexity.

It is possible that the blocks constructed by $\ex$ still cannot provide sufficiently strong constraints for derandomized local search. We will prove that in this case, $\pa_2$ is fast enough to solve $F$. $\pa_2$ enumerates particular assignments according to the result of $\ex$. Once these assignments are applied, the formula will be reduced to a $2$-CNF, which can be solved in polynomial time. We will prove that there would not be too many such assignments.

The most significant analytical difficulty is quantifying the gain of a block. We have simplified the analysis of chains from Liu's computing a linear programming to an inductive argument. However, blocks are harder because they are non-linear structures. In a block, a clause may connect several chains, and the restrictions created by different clauses can interact rather than contribute independently. Handling these interactions is the main analytical challenge of our work. We begin with the inductive bounds for the chains within a block, then examine how the additional clauses share variables with them. This gives a bound for every block produced by the algorithm without enumerating all blocks or computing each one’s exact characteristic value like the analysis of Liu.

We also generalize our method to the case where $k
\ge 4$. Our $3$-SAT algorithm balances two possibilities: either
the blocks make local search efficient, or partial
assignment reduces the formula to a $2$-CNF. For $k\ge4$, the second possibility requires a further decision. 
The question is how much work to do through
assignment and how much to leave to the residual solver.
The choices are to branch on the original short chains, to assign the
selected blocks make different choices, or to do a derandomized local search that uses
the same structure to solve the formula directly.
We compare all three complete costs, allowing a saving
in one procedure to compensate when another is expensive.
This extends the tradeoff behind our $3$-SAT improvement
to general $k$-SAT and yields the new bounds
reported in Table~\ref{tab:ksat-comparison}.

\begin{table}[ht]
\centering
\resizebox{\linewidth}{!}{%
\begin{tabular}{|c|c|c|c|c|c|c|c|}
\hline
$k$ & Our result
& \shortstack{Liu\\\cite{liu2018ksat}}
& \shortstack{Makino et al.\\
  \cite{DBLP:journals/algorithmica/MakinoTY13}}
& \shortstack{Moser--Scheder\\\cite{moser2011full}}
& \shortstack{Dantsin et al.\\\cite{dantsin2002deterministic}}
& \shortstack{PPSZ/Hertli\\(randomized)\\
  \cite{DBLP:journals/jacm/PaturiPSZ05,hertli20143}}
& \shortstack{Scheder\\(randomized)\\\cite{scheder2024ppsz}} \\
\hline\hline
3 & \textbf{1.327360} & 1.327924 & 1.330300 & 1.333334 & 1.481000
  & 1.307032 & $b_3-10^{-7}$ \\
\hline
4 & \textbf{1.498235} & 1.498570 & -- & 1.500001 & 1.600000
  & 1.468984 & $b_4-10^{-19}$ \\
\hline
5 & \textbf{1.599345} & 1.599451 & -- & 1.600001 & 1.666667
  & 1.569427 & $b_5-10^{-20}$ \\
\hline
6 & \textbf{1.666421} & 1.666455 & -- & 1.666667 & 1.714286
  & 1.637874 & $b_6-10^{-21}$ \\
\hline
\end{tabular}%
}
\caption{Exponential-base upper bounds for general $k$-SAT,
with polynomial factors suppressed.
All algorithms are deterministic except those in the last two columns.
In the Scheder column, $b_k$ is the unrounded classical PPSZ/Hertli
base, whose upward-rounded value appears in the preceding column.
The subtracted quantities are conservative improvements computed here
from Scheder's analysis and the Scheder--Steinberger lifting theorem
\cite{scheder2024ppsz,scheder2022ppszfull,
schedersteinberger2024generalppsz}.
}
\label{tab:ksat-comparison}
\end{table}



\section{Preliminaries}

In this section, we first give the concepts that are used in our algorithm, then present Liu's algorithms, which are still used in our algorithm.

\subsection{Notations}

\paragraph{Original Form of Formula.}
The original input $3$-CNF is denoted by $\mathcal F$.

\paragraph{Empty Clause.}
An empty clause is denoted by $\bot$ and indicates unsatisfiability.

\paragraph{Forms of Clauses.}
For every $C \in F$, let $C^F$ denote the clause $C$ in $F$ and $C^{F|\alpha} \in F|\alpha$ the new clause derived from $C$ by assigning variables according to $\alpha$, and $C^{\mathcal{F}}$ is called the \emph{original form} of clause $C$.

\paragraph{Set of $k$-clauses.}
For a CNF $F$, let $F_k:=\{C|C\in F,|C|=k\}$. $F_k$ denotes the set of all $k$-clauses in $F$. For a $3$-CNF $F$, $F = F_3 \mathbin{\dot\cup} F_2 \mathbin{\dot\cup} F_1$, while $F_2$ and $F_1$ may be empty.

\paragraph{Set of Variables.}
For any literal, clause, or any nested family thereof $\mathcal C$, let $V(\mathcal C)$ denote the set of all variables occurring in $\mathcal C$.

\paragraph{Applying a Partial Assignment.}
For a CNF $F$ and a partial assignment $\alpha$, let
$F|\alpha
:= \{ C \setminus \{ l \in C|\alpha(l)=0 \} | C \in F,\ \nexists l \in C \text{ with } \alpha(l)=1 \}$.
$F|\alpha$ denotes the formula obtained from applying the partial assignment $\alpha$ to $F$ and simplifying.

\paragraph{Unit Propagation.}
Let $F^{\mathrm{UP}}$ be the CNF derived by running \emph{Unit Propagation} on $F$ until there is no $1$-clause in $F$, and we write $F|_\alpha^{\mathrm{UP}}$ as shorthand for $(F|\alpha)^{\mathrm{UP}}$.
Clearly $F$ is satisfiable iff $F^{\mathrm{UP}}$ is satisfiable, and $\mathrm{UP}$ runs in polynomial time.

\paragraph{Ternary-to-Binary Clauses under UP.}
For a 3-CNF $F$ and a partial assignment $\alpha$, let $F_{3\to2}^\alpha:=\{C | C \in (F|_\alpha^{\mathrm{UP}})_2, C^F \in F_3\}$. $F_{3\to2}$ denotes the set of $2$-clauses that are originally $3$-clauses before apply assignment $\alpha$ and do unit propagation.

\paragraph{Touch.}
If two clauses or any nested families thereof $\mathcal C_1$ and $\mathcal C_2$ satisfies $V(\mathcal C_1)\cap V(\mathcal C_2)\neq\varnothing$, then we say $\mathcal C_1$ touches $\mathcal C_2$ or reflexively $\mathcal C_2$ touches $\mathcal C_1$.

\paragraph{Time Complexity.}
Let $O^*(f(n))$ denote $\operatorname{poly}(n)\cdot f(n)$ and $\mathcal O(f(n))$ denote $2^{o(n)}\cdot f(n)$.

\subsection{Preliminaries for Chains and Blocks}

\begin{definition}[Chain]
\label{def:chain}
Given integers $k\ge3$ and $\tau\ge1$, a $\tau$-\emph{chain} $\mathcal{S}^{(k)}$ is a sequence of $\tau$ $k$-clauses $\langle C_1, \dots, C_{\tau} \rangle$ such that $\forall i, j \in [\tau]$, $V(C_i) \cap V(C_j) = \varnothing$ iff $|i - j | > 1$.
\end{definition}

In other words, each clause in chain only and must overlap with the clauses next to it (if exist). If the context is clear, we will use $\mathcal{S}$, $\tau$-chain or simply chain for short.

\begin{definition}[Overlap Cases of Clause Sequences]
\label{def:overlap}
Let $C_i$ and $C_{i+1}$ be two consecutive clauses in a clause sequence of 3-clauses, and $V(C)$ denote the set of variables appearing in $C$. The overlap between them is classified into one of the following four types:

\begin{center}
\begin{tabular}{c|c}
\textbf{Name} & \textbf{Condition} \\ \hline
\text{independent} & $V(C_i) \cap V(C_{i+1}) = \varnothing$ \\[1.2ex]
\text{negative} & $V(C_i) \cap V(C_{i+1}) = \{x\}$ and $x$ occurs with opposite signs \\[1.2ex]
\text{positive} & $V(C_i) \cap V(C_{i+1}) = \{x\}$ and $x$ occurs with the same sign \\[1.2ex]
\text{two-negative} & $V(C_i) \cap V(C_{i+1}) = \{x, y\}$ and both $x,y$ occur with opposite signs
\end{tabular}
\end{center}

\end{definition}

No other overlap types occur in the branching procedure according to \cite{liu2018ksat}.

\begin{definition}[Encoding of Clause Sequences]
\label{def:to_string}
Let $\Gamma:=\{\verb"*",\verb"n",\verb"p",\verb"t"\}$. For a clause sequence $\mathcal C=\langle C_1,\dots,C_m\rangle$, define the encoding function $\zeta:\mathcal{C}\mapsto\Gamma^{|\mathcal{C}|}$ by $\zeta(\mathcal C)_m=\verb"*"$, and for $i \in [m-1]$,
\[
\zeta(\mathcal C)_i :=
\begin{cases}
\verb"*", &C_i\text{ and }C_{i+1}\text{ have an independent overlap},\\
\verb"n", &C_i\text{ and }C_{i+1}\text{ have a negative overlap},\\
\verb"p", &C_i\text{ and }C_{i+1}\text{ have a positive overlap},\\
\verb"t", &C_i\text{ and }C_{i+1}\text{ have a two-negative overlap}.
\end{cases}
\]
\end{definition}

Chains are key to the DLS speedup in the original algorithm by Liu \cite{liu2018ksat}. Naturally, one might expect more complex structures of clauses can better serves the purpose. We call these structures as \emph{components} and present one such structure here.
\begin{definition}[Block]
\label{def:block}
A $\tau$-\emph{block} $\mathcal B^{(k)}=(\mathcal C,\mathcal S)$ is a set of $\tau$ $k$-clauses that admits a partition into a clause set $\mathcal C=\{C_1,\ldots,C_u\}$ and a chain set $\mathcal D=\{\mathcal S_1,\ldots,\mathcal S_v\}$, such that:
\begin{enumerate}
    \item $V(\mathcal S_i)\cap V(\mathcal S_j)=\varnothing$ for all $i\ne j$.
    \item $V(\mathcal S_i)\cap V(\mathcal C)\ne\varnothing$ for all $i$.
\end{enumerate}
The clauses in $\mathcal C$ are called the \emph{center
clauses}, and the chains in $\mathcal D$ are called the \emph{leaf chains}.
\end{definition}

In other words, each clause in a block $\mathcal B^{(k)}$ must overlap with at least $1$ clause in $\mathcal B^{(k)}$. If the context is clear, we will use $\mathcal B$, $\tau$-block or simply block for short. The smallest possible block consists of only a single clause $C$ and $\mathcal B=(\{C\},\varnothing)$.

\begin{definition}[Isomorphic of Components]
\label{def:isomorphic_block}
Two components $\mathcal K_1$ and $\mathcal K_2$ are \emph{isomorphic} iff $\exists$ singled literal bijection $\phi$, $\{\phi(C)|C\in\mathcal K_1\}=\mathcal K_2$ and we say that $\mathcal K_1$ and $\mathcal K_2$ have the same type.
\end{definition}

If both $\mathcal K_1$ and $\mathcal K_2$ are chains, it is obvious that this is equivalent to $\zeta(\mathcal K_1)=\zeta(\mathcal K_2)$. We use $\mathcal S_i$ to denote the chain of type-$i$, and $\mathcal B_i$ to denote the block of type-$i$ if there are only finite type of chains or blocks.

\begin{definition}[Generalized Instance]
\label{def:generalized_instance}
A collection of components $\mathcal{I}$ is called an \emph{instance} if $\forall \mathcal K,\mathcal K'\in\mathcal{I},V(\mathcal{K})\cap V(\mathcal{K}')\neq\varnothing \text{ iff } \mathcal K=\mathcal K'$.
\end{definition}

That is to say components in an instance are mutually independent. When each components $\mathcal K\in\mathcal I$ is a chain, we cover the original definition of instance.

\begin{definition}[Solution Space]
\label{def:solution_space}
Given a component $\mathcal{K}$, define its \emph{solution space} as 
\[
A(\mathcal{K}):=\{\alpha\in\{0,1\}^{|V(\mathcal{K})|}|\forall C\in\mathcal{K},\alpha\vDash C\}.
\]

And given an instance $\mathcal I$, define its \emph{solution space} as 
\[
A(\mathcal{I}):=\prod_{\mathcal K\in\mathcal I}A(\mathcal K).
\]
\end{definition}

\begin{definition}[Generalized Characteristic System]
\label{def:generalized_characteristic_system}
Given a component $\mathcal K$, let $A:=A(\mathcal K)$, and let
$d(a,a^*)$ denote the Hamming distance between $a,a^*\in A$. Define the
linear system $\mathrm{LP}_A$ with variables $\lambda_{\mathcal K}\in\mathbb{R}$ and
$\pi_{\mathcal K}(a)\in[0,1]$ for each $a\in A$ as follows:
\[
\begin{aligned}
&\sum_{a \in A}\pi_{\mathcal K}(a)=1,\\
&\lambda_{\mathcal K}=\sum_{a\in A}\pi_{\mathcal K}(a)\cdot{(\frac{1}{k-1})}^{d(a,a^*)} & \text{for all } a^* \in A, \\
& \pi_{\mathcal K}(a) \ge 0 & \text{for all } a \in A.
\end{aligned}
\]
If this system is feasible, its solution is unique. We call its coordinates
the \emph{characteristic value} $\lambda_{\mathcal K}$ and the \emph{characteristic
distribution} $\pi_{\mathcal K}$ of $\mathcal K$, respectively.
\end{definition}

Whenever the characteristic system is feasible, $\lambda_{\mathcal K}\in(0,1)$.
$\lambda_{\mathcal K}$ is an algebraic invariant of the solution space $A(\mathcal{K})$, measuring the efficiency of covering $A(\mathcal{K})$ by Hamming balls: the larger $\lambda_{\mathcal K}$ is, the more efficient the covering. The distribution $\pi_{\mathcal K}(a)$ weights an assignment $a$ according to its isolation from the rest of $A(\mathcal{K})$: the larger $\pi_{\mathcal K}(a)$ is, the more isolated $a$ is from the other points.

\subsection{The Derandomized Local Search}

We use Liu's derandomized local search (\dls)~\cite{liu2018ksat}, as Algorithm~\ref{alg:dls}.

\begin{algorithm}[H]
\small
\setlength{\linewidth}{\hsize}
\caption{Derandomized Local Search: \textsf{DLS}}
\label{alg:dls}
\begin{algorithmic}[1]
\REQUIRE $3$-CNF $F$, instance $\mathcal I$
\ENSURE a satisfying assignment or $\textsf{Unsatisfiable}$
\STATE construct covering code $\mathfrak C$ for Hamming space $H(F,\mathcal I)$
\FOR{$\alpha\in\mathfrak C$}
    \IF{$\textsf{searchball-fast}(F,\alpha,r)$ finds a satisfying assignment $\alpha^*$ for $F$}
        \RETURN $\alpha^*$ 
    \ENDIF
\ENDFOR
\end{algorithmic}
\end{algorithm}

Given a $3$-CNF $F$ and an instance $\mathcal I$, \dls{} constructs
a generalized covering code $\{\mathfrak C(r)\}_r$ for $H(F,\mathcal I)$,
the product of the solution spaces in $\mathcal I$ and the Boolean
cube on the remaining variables. For each radius $r$ and each
$\alpha\in\mathfrak C(r)$, it uses \textsf{searchball-fast} to search
the Hamming ball of radius $r$ centered at $\alpha$. It returns the
first satisfying assignment found, or \textsf{Unsatisfiable} after
all balls have been searched. This is complete because every
satisfying assignment of $F$ belongs to $H(F,\mathcal I)$.

\begin{lemma}[Running Time of \dls~\cite{liu2018ksat}]
\label{lem:dlstime}
Given a $3$-CNF $F$ on $n$ variables and a chain instance $\mathcal I$,
\dls{} runs in time
\[
T_{\dls}=\mathcal O\!\left(
    \left(\frac{4}{3}\right)^{n'}\prod_i\lambda_i^{-\nu_i}
\right),
\]
where $n'=n-|V(\mathcal I)|$, $\nu_i$ is the number of chains of type
$i$, and $\lambda_i$ is their characteristic value.
\end{lemma}

\subsection{The Branching Algorithm for 3-SAT}

We present the branching algorithm ($\br$) of Liu in this Subsection. This is the key to construct chains.

\subsubsection{Simplification Procedure}

The simplification relies on the following two lemmas.

\begin{lemma}[Autark \cite{liu2018ksat}]
\label{lem:autark}
For a 3-CNF $F$ and a partial assignment $\alpha$, if $\bot\notin F|_\alpha^{\mathrm{UP}}$ and $F_{3\to2}^\alpha=\varnothing$, then $F$ is satisfiable iff $F|_\alpha^{\mathrm{UP}}$ is satisfiable. Such an assignment $\alpha$ is called an autark.
\end{lemma}

\begin{lemma}[Binary Clause Reduction \cite{liu2018ksat}]
\label{lem:binary_deduction}
For a 3-CNF $F$ and a binary clause $B \in F_2$, if there exist a literal $l \in B$ and a clause $C \in F_{3\to2}^{l=1}$ such that $C \cap B \neq \varnothing$, then $F$ is satisfiable iff $F\setminus\{C^F\}\cup\{C\}$
is satisfiable.
\end{lemma}

As a result, the simplification function $\mathcal P$ is as follows.

\begin{definition}[Simplification Function $\mathcal{P}$]
\label{def:procedure_p}
For a 3-CNF $F$, define
\[
\mathcal{P}(F) :=
\begin{cases}
\mathcal{P}(F|_{l=1}^{\mathrm{UP}}),&\text{if }\exists B\in F_2,\ \exists l\in B\text{ such that } \bot\notin F|_{l=1}^{\mathrm{UP}}\text{ and }F_{3\to2}^{l=1}=\varnothing; \\[1.2ex]
\mathcal{P}(F\setminus\{C^F\}\cup\{C\}),
& \text{else if } \exists B \in F_2,\ \exists l \in B,\ \exists C \in F_{3\to2}^{l=1} \text{ with } C \cap B \neq \varnothing; \\[1.2ex]
F,
& \text{otherwise.}
\end{cases}
\]
\end{definition}

By definition, $\mathcal{P}(F)$ is a fixed point of $\mathcal{P}$, i.e., $\mathcal{P}(\mathcal{P}(F)) = \mathcal{P}(F)$. $\mathcal{P}(F)$ is satisfiable iff $F$ is satisfiable by Lemma~\ref{lem:autark} and Lemma~\ref{lem:binary_deduction}.

Clearly $\mathcal P$ is computable in polynomial time: each recursive application strictly reduces the formula size, and the conditions are checkable in polynomial time. $\mathcal P$ is deterministic if we traverse the clauses in a fixed order.

\begin{lemma}[No-Overlap Property \cite{liu2018ksat}]
\label{col:no_overlap}
Given a 3-CNF $F$, for all $B\in(\mathcal{P}(F))_2$ and $l\in B$, we have 
\[
C\cap(B\setminus\{l\})=\varnothing\quad \text{for every } C \in (\mathcal{P}(F))_{3\to2}^{l=1}.
\]
\end{lemma}

\subsection{The Branching Procedure}

\subsubsection{The Branching Algorithm}

We now present our branching algorithm \textsf{BR} for $3$-SAT. At the beginning, the inputs $\mathcal{C}$, $C_{\text{prev}}$ and $\alpha_{\text{prev}}$ are all $\varnothing$.
For a clause sequence $\mathcal C$ and a clause $D$, write
$\mathcal C\mathbin{\|}D$ for the sequence obtained by appending $D$ to
$\mathcal C$.

\begin{algorithm}[H]
\small
\setlength{\linewidth}{\hsize}
\caption{Branching Algorithm \textsf{BR} for $3$-SAT}
\label{alg:branch}
\begin{algorithmic}[1]
\REQUIRE $3$-CNF $F_{\mathrm{parent}}$, clause sequence $\mathcal{C}$, partial assignment $\alpha_{\text{prev}}$
\ENSURE a satisfying assignment $\alpha$ or \textsf{Unsatisfiable} or an instance $\mathcal{I}$
\STATE $F \leftarrow \mathcal P(F_{\mathrm{parent}})$
\IF{$F=\varnothing$}
    \RETURN the current assignment
\ELSIF {$\bot \in F$}
    \RETURN \textsf{Unsatisfiable} \label{line_unsat1}
\ELSIF {$\log{b(\mathcal C)}\ge n\log c$}
    \STATE $\mathcal{I} \leftarrow \textsf{TransInstance}(\mathcal C)$
    \STATE stop the recursion and $\textbf{return}$ $\mathcal{I}$
\ELSIF {$F_3=\varnothing$}
    \STATE solve $F$ in polynomial time
    \IF{$F$ is solved}
        \STATE stop the recursion and $\textbf{return}$ the satisfying assignment
    \ELSE
        \RETURN $\textsf{Unsatisfiable}$
    \ENDIF
\ELSE
    \STATE $l\leftarrow$ and arbitrary literal that is set to $1$ in $\alpha_\text{prev}$
    \STATE $C\leftarrow$ and arbitrary clause in $F_{3\to2}^{l=1}$
    \FOR {$\alpha\in A(C)$}
        \STATE call $\textsf{BR}(F|_{\alpha}^{\mathrm{UP}},\mathcal{C}\mathbin{\|}C^\mathcal F,\alpha)$
    \ENDFOR
    \RETURN \textsf{Unsatisfiable}
\ENDIF
\end{algorithmic}
\end{algorithm}

We can always begin the branching by choosing an arbitrary $2$-clause and enumerating its satisfying assignments. If there is not $2$-clause, then branching on an arbitrary variable. If this produces a $2$-clause, then begin the branching, otherwise it is an autark, we fix it and branch on another variable. This costs polynomial time and add an extra factor of $2$ on the upper bound, which would not change it.

$\textsf{TransInstance}(\mathcal{C})$ constructs an instance $\mathcal{I}$ as follows: split $\zeta(\mathcal{C})$ at every occurrence of $\verb"*"$ in $\mathcal C$; each resulting segment forms a chain, and the set of all such chains is $\mathcal I$; then returns $\mathcal{I}$.

$\br$ is central to the construction of chains. At each recursive step, it selects an original $3$-clause that part of the assignment made at the parent node reduces to a $2$-clause, then branches over the satisfying assignments of that $2$-clause. The clause need not still appear in the current simplified formula: the full branch assignment may have simplified it, but it can still be used to define the branches. If $F$ is reduced to a $2$-CNF, then it is solved immediately in polynomial time. When the accumulated logarithmic branching cost reaches $n\log c$, where $c$ is the base of the overall running-time bound, $\br$ stops and converts the branching-clause history into a chain instance $\mathcal I_{\text{chain}}$. Correctness follows because every simplification preserves satisfiability and every explicit branch exhausts the relevant assignments.

\begin{lemma}[Structure of the recorded history]
\label{lem:recorded-history}
Every recorded original clause can intersect only its immediate predecessor.
Consequently, splitting the recorded sequence at independent adjacencies
produces an instance of variable-disjoint chains.
\end{lemma}

\begin{proof}
When a clause is selected, it is ternary in the saved parent formula. All
variables belonging to clauses earlier than the immediate predecessor have
already been assigned. A surviving ternary clause contains none of those
variables; otherwise it would have been satisfied and removed, or shortened.
The triggering assignment can shorten it through a variable of the immediate
predecessor, which is therefore its only possible earlier overlap. The claim
follows inductively along the recorded sequence.
\end{proof}

\subsubsection{Branching Costs}
\label{subsec:branch}

\begin{lemma}[Admissible overlap profiles]
\label{lem:admissible-profiles}
After an independent, negative, or positive overlap, any of $n,p,t$ may occur
next. Within a chain, a two-negative overlap can be followed only by $n$.
Equivalently, the substrings $tp$ and $tt$ never occur.
\end{lemma}

\begin{proof}
Normalize the two clauses in a two-negative overlap to
$l_0\lor l_1\lor l_2$ and $\bar l_1\lor\bar l_2\lor l_3$. In the branches
with $l_3=0$, the extra binary split begins a new independent chain. In the
remaining branches, $l_3=1$ is the trigger for the next selection. A clause
containing $l_3$ positively is then satisfied and removed, so a nonindependent
successor can share only $\bar l_3$ with the current clause. Since the older
variables have already been assigned, a second two-variable overlap is also
impossible. Thus only a negative overlap can continue the same chain.
\end{proof}

Taking the maximum branch number $3$ for the \textbf{n} cases gives the local
branching costs below.

\begin{table}[htbp]
\centering
\begin{tabular}{c|c|c|c|c}
\textbf{Overlap Case} & \textsf{*} & \textsf{n} & \textsf{p} & \textsf{t} \\ \hline
\textbf{Branch Number} & $3$ & $3$ & $2$ & $7/3$
\end{tabular}
\caption{Branching cost contributed by each overlap type \cite{liu2018ksat}.}
\label{tab:branchnumber}
\end{table}

For a chain $\mathcal{S}$, its branch number is given by
\[
b(\mathcal{S})=2^{|\zeta(\mathcal S)|_\textsf{p}}\cdot3^{|\zeta(\mathcal S)|_\textsf{n}+1}\cdot(\frac{7}{3})^{|\zeta(\mathcal S)|_\textsf{t}}
\]
where $|\zeta(\mathcal{S})|_s$ denotes the number of occurrences of $s$ in $\mathcal{S}$. The symbol \textsf{*} is omitted from the formula because $|\zeta(\mathcal S)|_\textsf{*}=1$ for every chain.

For a clause sequence $\mathcal{C}$, let $\mathcal{I}(\mathcal{C})$ denote the
instance obtained from $\mathcal{C}$. Its branching cost is
\begin{equation}
\label{eq:branch_of_c_1}
b(\mathcal{C})=2^{|\zeta(\mathcal C)|_p}\cdot3^{|\zeta(\mathcal C)|_n+|\zeta(\mathcal C)|_*}\cdot(\frac{7}{3})^{|\zeta(\mathcal C)|_t}=\prod_{\mathcal{S}\in\mathcal{I}(\mathcal{C})}b(\mathcal{S}).
\end{equation}

\begin{corollary}[Running Time of \textsf{BR}]
\label{cor:brtime}
Let $\mathcal{C}$ be the clause sequence of \textsf{BR} at running time $T$. Then
\[
T\leq O^*\left(\prod_{\mathcal S\in\mathcal I(\mathcal C)}
b(\mathcal S)\right)=O^*(b(\mathcal{C})).
\]
\end{corollary}
\begin{proof}
The claim follows from Table~\ref{tab:branchnumber} and
Equation~\eqref{eq:branch_of_c_1}.
\end{proof}
\section{Warmup: A Minor Improvement}

In this section, we illustrate how the structural information we gain from the branching procedure can be used to speed up the solution of the problem.

The achievement ca,n be achieved by using the simplest implementation of our method, in particular reducing the base from $c\approx1.32793$ to $c^*\approx c-2\times 10^{-14}$. Although the improvement is small, it is sufficient to show that out method is effective.

\begin{lemma}
\label{lem:upperbound_simple}
There exists a deterministic algorithm for $3$-SAT that runs in time $\mathcal O((c^*)^n)$. \footnote{$\mathcal O$ suppress subexponential factors, i.e. $\mathcal O(f(n)) = O(2^{o(n)}f(n))$.}
\end{lemma}

\subsection{The Simple Version Algorithm}

In Liu's original algorithm, if $\br$ does not solve $F$ within the allowed branching steps, we use its output $\mathcal I_\text{chain}$ to narrow the search space of $\dls$. In Algorithm \ref{alg:framework_simple}, if $\dls$ cannot solve $F$ within the allowed time, we construct an instance of blocks $\mathcal I_\text{block}$ that provide stronger constraints. Then we determine whether to solve $F$ by $\pa_2$ or $\dls$ according to the size of $\mathcal I_\text{block}$.

\begin{algorithm}[H]
\small
\setlength{\linewidth}{\hsize}
\caption{Simple Version of Algorithmic Framework for $3$-SAT}
\label{alg:framework_simple}
\begin{algorithmic}[1]
\REQUIRE $3$-CNF $F$
\ENSURE a satisfying assignment $\alpha$ or \textsf{Unsatisfiable}
\STATE $\br(F,\varnothing)$ either solves $F$ or returns a chain instance $\mathcal I_\text{chain}$
\IF{$F$ is not solved}
    \IF{$F$ can be solved by $\dls(F,\mathcal I_\text{chain})$ in allowed time}
        \STATE call $\dls(F,\mathcal I_\text{chain})$
    \ELSE
        \STATE $\mathcal I_\text{1-chain}\leftarrow\{\mathcal S|(\mathcal S\in\mathcal I_\text{chain})\land(\mathcal S\text{ is a $1$-chain})\}$
        \STATE $\mathcal I_\text{block}\leftarrow\textsf{Simple-}\ex(F,\mathcal I_\text{1-chain})$
        \IF{$|\mathcal I_\text{block}|\le\sigma n$}
            \STATE call $\pa_2(F,\mathcal I_\text{block},\mathcal I_\text{1-chain})$
        \ELSE
            \STATE call $\dls(F,\mathcal I_\text{block}\cup\mathcal I_\text{1-chain})$        
        \ENDIF
    \ENDIF
    
\ENDIF
\end{algorithmic}
\end{algorithm}

Whether $F$ can be solved by $\dls$ in allowed time is judged by whether its running time upper bound give by Lemma~\ref{lem:dlstime} is lower than $(c^*)^n$.

The subroutines are given below.

\begin{algorithm}[H]
\small
\setlength{\linewidth}{\hsize}
\caption{\textsf{Simple-}\ex}
\label{alg:expansion_simple}
\begin{algorithmic}[1]
\REQUIRE $3$-CNF $F$, instance of chains $\mathcal I_\text{1-chain}$
\ENSURE an instance of blocks $\mathcal I_\text{block}$
\STATE $\mathcal I_\text{block}\leftarrow\varnothing$
\FOR{$C\in\{C|\forall\mathcal S\in\mathcal I_\text{1-chain},C\notin\mathcal S\}$}
    \STATE $\mathcal B\leftarrow\{C\}\cup\{C_0|\exists\mathcal S\in\mathcal I_\text{1-chain},\mathcal S=\langle C_0\rangle\text{ and }V(C)\cap V(C_0)\neq\varnothing\}$
    \IF{$V(\mathcal B)\cap V(\mathcal I_\text{block})=\varnothing$}
        \STATE $\mathcal I_\text{block}\leftarrow\mathcal I_\text{block}\cup\{\mathcal B\}$
        \STATE Remove all chains in $\mathcal B$ from $\mathcal I_\text{1-chain}$
    \ENDIF
\ENDFOR
\RETURN $\mathcal I_\text{block}$
\end{algorithmic}
\end{algorithm}

Algorithm~\ref{alg:expansion_simple} is a simple version of the block construction. To make it easy to analyze, we only use the $1$-chains here. The algorithm greedily scans all clauses that are not in any $1$-chains. If it is disjoint from any existing blocks, then treat the clause and all its touched $1$-chains as a block. It also removes all chains that are used to construct a block from $\mathcal I_\text{chain}$, so the constrain of $\mathcal I_\text{1-chain}$ and $\mathcal I_\text{block}$ would nor conflict in $\dls$.

\begin{algorithm}[H]
\small
\setlength{\linewidth}{\hsize}
\caption{Reduce the Formula to a $2$-CNF: \textsf{Simple-PA2}}
\label{alg:pa2_simple}
\begin{algorithmic}[1]
\REQUIRE $3$-CNF $F$, instance of chains $\mathcal I_\text{1-chain}$, instance of blocks $\mathcal I_\text{block}$
\ENSURE a satisfying assignment or $\textsf{Unsatisfiable}$
\STATE $A\leftarrow A(\mathcal I_\text{block})\times \{0,1\}^t$, where $t={|\{\mathcal S|(\mathcal S\in\mathcal I_\text{1-chain})\land(\forall \mathcal B\in\mathcal I_\text{block},\mathcal S\nsubseteq\mathcal B)\}}$
\FOR{$\alpha\in A$}
    \STATE solve the $2$-CNF $F|_\alpha^{\text{UP}}$ in polynomial time
    \IF{$F$ is solved}
        \RETURN the satisfying assignment
    \ENDIF
\ENDFOR
\RETURN $\textsf{Unsatisfiable}$
\end{algorithmic}
\end{algorithm}

Algorithm~\ref{alg:pa2_simple} is the simple version of $\pa_2$. It enumerates all satisfying assignments of all blocks, and the $0/1$ assignment of an arbitrary variable of all $1$-chains that are not in any blocks. This solving terminal ensures that our algorithm would not perform too poorly when $\textsf{Simple-}\ex$ when there are not enough blocks to constrain the search space of $\dls$.

Lemma~\ref{lem:transto2cnf} shows that applying any of the enumerated assignments reduces $F$ to a $2$-CNF.

\begin{lemma}
\label{lem:transto2cnf}
In \textsf{Simple-Reduce2SAT}, $\forall\alpha\in A$, $(F|_\alpha^{\text{UP}})_3=\varnothing$. 
\end{lemma}

\begin{proof}

We show that $A$ assigns al least $1$ variable in $C$. Let $U=\bigcup_{\mathcal B\in\mathcal I_\text{block}}T(\mathcal B)$.

\begin{itemize}
    \item If $\langle C\rangle\in\mathcal I_\text{1-chain}$.
    If $C\in U$, $A$ assigns all variables in $C$. Otherwise, $A$ assigns an arbitrary variable $x\in V(C)$ by definition.
    \item If $\langle C\rangle\notin\mathcal I_\text{1-chain}$. Let $\mathcal B_C$ be the block in Algorithm~\ref{alg:expansion_simple} whose center clause is $C$. If $\mathcal B_C\in\mathcal I_\text{block}$, then $A$ assigns all variables in $C$ by definition. Otherwise, it must be $V(\mathcal B_C)\cap V(\mathcal I_\text{block})\neq\varnothing$. This occurs either immediately $V(C)\cap V(\mathcal I_\text{block})\neq\varnothing$, or $\exists C'\in T(\mathcal B_C),V(C')\cap V(\mathcal I_\text{block})\neq\varnothing$ and $V(C)\cap V(C')\neq\varnothing$. For the later one, according to the disjointness of the instance $\mathcal I_\text{1-chain}$ it must be $\exists\mathcal B'\in\mathcal I_\text{block},C'\in T(\mathcal B')$, thus we also have $V(C)\cap V(\mathcal I_\text{block})\neq\varnothing$. Then either case $A$ assigns at least $1$ variable in $C$.
\end{itemize}

Therefore, applying the assignment $\alpha$ either satisfies each
original $3$-clause or reduces its size to at most two.
\end{proof}

\subsection{A Rough Upper Bound Analysis}

In this subsection we will use a simple method to prove that the upper bound of the simple version algorithm $(c^*)^n$ is strictly better than Liu's original one $c^n$. Assume that $\br$ follows Liu's rule-$2$ truncation \cite{liu2018ksat}.

There is only $1$ center clause, and all leaf chains are $1$-chains in the simplified algorithm. We write $t(\mathcal B)$ as the number of leaf chains and $v(\mathcal B)$ as the number of the variables that the center clause does not share with any leaf chains. Then we have the following Lemma~\ref{lem:range_block}.

\begin{lemma}[The Range of $t(\mathcal B)$ and $v(\mathcal B)$]
\label{lem:range_block}
For a block $\mathcal B^{(k)}$, $t(\mathcal B^{(k)})+v(\mathcal B^{(k)})\le k$.
\end{lemma}

\begin{proof}
Each leaf clauses at least share $1$ variable with $C(\mathcal B)$. Then the conclusion follows.
\end{proof}

\begin{lemma}
\textsf{Simple-Reduce2SAT} runs in time $T_\textsf{S-RD}\le \mathcal O(2^{(\frac{\log{c^*}}{\log{3}}+6\sigma)n}).$
\end{lemma}

\begin{proof}
It is obvious that $T_\textsf{S-RD}=\mathcal O(|A|)$ where $A$ is set of assignments enumerated by $\textsf{Simple-}\pa_2$.

By the terminate condition of $\br$,
\[
\nu_1\log{3}\le\sum_{i\in[\theta]}\nu_i\log{b_i}=n\log{c^*}+O(1),
\]
where $\theta$ is the number of chain types.

Then we know that the number of $1$-chains in $\mathcal I_\text{chain}$ $\nu_1$ satisfies
\[
\nu_1\le \frac{n\log{c^*}}{\log{3}}+O(1).
\]

Let $T=\sum_{\mathcal B\in\mathcal I_\text{block}}t(\mathcal B)$ and $V=\sum_{\mathcal B\in\mathcal I_\text{block}}v(\mathcal B)$. By construction, each block $\mathcal B$ consists of a
center clause $C$ and $t(\mathcal B)$ pairwise
variable-disjoint singleton clauses.
Each singleton contains at least one distinct variable of
$C$, while $v(\mathcal B)$ counts the variables of $C$
outside these singletons. Therefore,
\[
t(\mathcal B)+v(\mathcal B)\le |V(C)|\le3.
\]
Summing over the selected blocks gives
\[
T+V
=\sum_{\mathcal B\in\mathcal I_{\mathrm{block}}}
  \bigl(t(\mathcal B)+v(\mathcal B)\bigr)
\le3|\mathcal I_{\mathrm{block}}|.
\]

For every block $\mathcal B$, any of its satisfying assignments should at least satisfy all of its leaf clauses, so we have $|A(\mathcal B)|\le 7^{t(\mathcal B)}\cdot 2^{v(\mathcal B)}$. So, the number of partial assignments enumerated in $\textsf{Simple-}\pa_2$ satisfies
\[
A=2^{\nu_1-T}\cdot\prod_{\mathcal B\in\mathcal I_\text{block}}|A(\mathcal B)|\le2^{\nu_1-T}\cdot7^T\cdot 2^V\le2^{\nu_1+2T+V}\le2^{\nu_1+6|\mathcal I_\text{block}|}.
\]

According to the condition of choosing $\textsf{Simple-}\pa_2$, $|\mathcal I_\text{block}|\le\sigma n$, so
\[
A\le2^{\nu_1+6\sigma n}\le2^{(\frac{\log{c^*}}{\log{3}}+6\sigma)n+o(n)}.
\]

Then we have
\[
T_\textsf{S-RD}\le \mathcal O(2^{(\frac{\log{c^*}}{\log{3}}+6\sigma)n}).
\]
\end{proof}


\begin{lemma}[Uniform block gain]
\label{lem:warmup-block-gap}
Every selected block satisfies
\begin{equation}
\label{eq:warmup-block-gap}
 \lambda_{\mathcal B}
 \ge(1+\varepsilon)
       (3/7)^{t(\mathcal B)}(3/4)^{v(\mathcal B)},
\end{equation}
where
\[
 \varepsilon=\frac{1}{2^3\cdot21^6}.
\]
\end{lemma}

\begin{proof}
Fix a block, and abbreviate $t=t(\mathcal B)$, $v=v(\mathcal B)$,
and $d=3t+v$. Complement coordinates so that every touched source clause
is positive. Let $A_0=\{0,1\}^3\setminus\{000\}$, and define
\[
 A'=A_0^t\times\{0,1\}^v,
 \qquad A=A(\mathcal B)\subsetneq A'.
\]
The inclusion is strict because the witness is a proper clause distinct
from every source, since every assignment with the last $v$ bit being $0$ is in $A'$ but not $A$. Also $A$ is nonempty: making one witness literal true is
compatible with satisfying all the disjoint source clauses.

For a word $x\in A_0$, let $|x|$ denote its number of one-bits. The
singleton characteristic distribution is
\[
 \pi_0(x)=
 \begin{cases}
  4/21,&|x|=1,\\
  2/21,&|x|=2,\\
  1/7,&|x|=3.
 \end{cases}
\]
Its characteristic value is $3/7$. A free bit has the uniform
distribution and characteristic value $3/4$. Their product distribution
$\pi'$ on $A'$ consequently has characteristic value
\[
 \lambda'=(3/7)^t(3/4)^v
\]
and satisfies
\[
 \pi'(x)\ge(2/21)^t(1/2)^v\ge(2/21)^3
 \qquad(x\in A'),
\]
because $t+v\le3$.

Let $K_{A'}$ be the matrix indexed by $A'$ with entry
$(K_{A'})_{x,y}=2^{-d_H(x,y)}$. Its product distribution satisfies
$K_{A'}\pi'=\lambda'\mathbf1$, where $\mathbf1$ is the all-ones
vector. The full $d$-bit matrix is the $d$-fold tensor product of
\[
 \begin{pmatrix}1&1/2\\1/2&1\end{pmatrix}.
\]
Its smallest eigenvalue is $2^{-d}$. Since $K_{A'}$ is a principal
submatrix and $d\le9$, its smallest eigenvalue is at least $2^{-9}$.

Let $\Delta(A)$ denote the set of probability distributions on $A$, i.e.
\[
\Delta(A)=\left\{r\to[0,1]:\sum_{x\in A}r(x)=1\right\}.
\]
Take any distribution $r\in\Delta(A)$ and extend it by zero to a vector
$\widetilde r$ on $A'$. Some removed word has $\widetilde r(x)=0$ and
$\pi'(x)\ge(2/21)^3$. Thus, with $\|\cdot\|_2$ denoting Euclidean norm,
\[
 \|\widetilde r-\pi'\|_2^2\ge(2/21)^6.
\]
Both distributions have total mass one. Expanding the quadratic form
therefore cancels the cross term and yields
\begin{align*}
 r^{\mathsf T}K_A r
 &=\lambda'
   +(\widetilde r-\pi')^{\mathsf T}K_{A'}(\widetilde r-\pi')\\
 &\ge\lambda'+2^{-9}(2/21)^6
 =\lambda'+\varepsilon,
\end{align*}
where $K_A$ is the corresponding matrix indexed by $A$.
Since $r^{\mathsf T}K_A r$ is an average of the entries of $K_A r$,
\[
 \max_{x\in A}(K_A r)_x\ge\lambda'+\varepsilon
 \qquad\text{for every }r\in\Delta(A).
\]
The finite minimax theorem and symmetry of $K_A$ now give
\[
 \lambda(A)
 =\min_{r\in\Delta(A)}\max_{x\in A}(K_A r)_x
 \ge\lambda'+\varepsilon
 \ge(1+\varepsilon)\lambda',
\]
where the final inequality uses $\lambda'\le1$.
\end{proof}

\begin{lemma}
\label{lem:original_dls_time}
For a given instance $\mathcal I$, $\dls$ runs in time $T_\dls=\mathcal O((\frac{4}{3})^{n_\mathrm{free}}\prod_{\mathcal K\in\mathcal I}\lambda_{\mathcal K}^{-1})$ where $n_\mathrm{free}=n-|V(\mathcal I)|$.
\end{lemma}

\begin{lemma}
\label{lem:num_1_chain}
Let $\nu_1=|\mathcal I_\text{1-chain}|$ before $\textsf{Simple-}\ex$ modifies it. If the upper bound of $\dls(F,\mathcal I_\text{chain})$ is greater than $(c^*)^n$, then we have
\[
\nu_1>\frac{(1+\kappa+\mu)\log{c^*}-\log{4/3}}{\mu\log{3}}n>0
\]
where $\kappa=\frac{|V(\mathcal S_1)|\log{4/3}+\log{\lambda_{\mathcal S_1}}}{\log{b(S_1)}}$ and $\mu=\min_{i\neq1}\{\frac{|V(\mathcal S_i)|\log{4/3}+\log{\lambda_{\mathcal S_i}}}{\log b(S_i)}-\kappa\}$, where $\mathcal S_i$ denotes type-$i$ chain under Liu's all possible types of chains under the rule-$2$ truncation \cite{liu2018ksat}.
\end{lemma}

\begin{proof}
Among all possible chains under Liu's rule-$2$, $\forall i\in[\theta],Q_1/\log{b_1}\le Q_i/\log{b_i}$, so we have $\mu>0$.

The logarithm running time of $\dls$ is $\log T_\dls=n\log(4/3)-\sum_{i\in[\theta]}Q_i$, then
\[
n\log(4/3)-\sum_{i\in[\theta]}\nu_iQ_i\ge n\log c^*.
\]

For all chains except $1$-chain we have $Q_i\ge(\kappa+\mu)\log b_i$ by the definition of $\mu$. Then we have
\[
n\log(4/3)-\sum_{i\in[\theta]\land i\neq1}\nu_i(\kappa+\mu)\log b_i-\kappa\nu_1\log b_1\ge n\log c^*.
\]

By the terminate condition of $\br$ we have $\sum_{i\in[\theta]}\nu_i\log b_i>n\log c^*$, so
\[
n\log(4/3)-(\kappa+\mu)n\log c^*+\mu\nu_1\log b_1>n\log c^*.
\]

Rearranging terms yields
\[
n\log(4/3)-\sum_{i\in[\theta]}\nu_iQ_i\ge n\log c^*.
\]
\end{proof}

\begin{lemma}
\label{lem:large-block-dls}
Let
\[
 \alpha=
 \frac{(1+\kappa+\mu)\log c^*-\log(4/3)}
      {\mu\log 3}.
\]
Suppose that the large-block branch invokes $\dls$ on the selected
blocks together with all untouched singleton chains. Then
\[
 T_\dls\le
 \mathcal O\!\left(
   \left[
     \frac43
     \left(\frac{63}{64}\right)^\alpha
     (1+\varepsilon)^{-\sigma}
   \right]^n
 \right).
\]
In particular, $T_\dls\le\mathcal O((c^*)^n)$ whenever
\[
 \log(4/3)
 -\alpha\log(64/63)
 -\sigma\log(1+\varepsilon)
 \le \log c^*.
\]
\end{lemma}

\begin{proof}
Let $m=|\mathcal I_{\mathrm{block}}|$. The condition selecting
this branch gives $m>\sigma n$. Define
\[
 k=\sum_{\mathcal B\in\mathcal I_{\mathrm{block}}}
          t(\mathcal B),
 \qquad
 q=\sum_{\mathcal B\in\mathcal I_{\mathrm{block}}}
          v(\mathcal B).
\]
Thus $k$ is the number of singleton sources covered by the blocks,
and $q$ is the number of covered variables outside those sources.

The selected blocks have disjoint variable supports. Together,
they contain $3k+q$ variables, while the $\nu_1-k$ untouched
singleton chains contain another $3(\nu_1-k)$ variables.
Hence the $\dls$ instance leaves $n-3\nu_1-q$ free variables.

Applying the \dls running-time bound and the singleton
characteristic value $3/7$ gives
\[
 T_\dls\le
 \mathcal O\!\left(
   \left(\frac43\right)^{n-3\nu_1-q}
   \left(\frac73\right)^{\nu_1-k}
   \prod_{\mathcal B\in\mathcal I_{\mathrm{block}}}
        \lambda_{\mathcal B}^{-1}
 \right).
\]
By Lemma~\ref{lem:warmup-block-gap},
\[
 \prod_{\mathcal B\in\mathcal I_{\mathrm{block}}}
        \lambda_{\mathcal B}^{-1}
 \le
 (1+\varepsilon)^{-m}
 \left(\frac73\right)^k
 \left(\frac43\right)^q.
\]
Substituting and simplifying yields
\begin{align*}
 T_\dls
 &\le
 \mathcal O\!\left(
   \left(\frac43\right)^{n-3\nu_1}
   \left(\frac73\right)^{\nu_1}
   (1+\varepsilon)^{-m}
 \right)\\
 &=\mathcal O\!\left(
   \left(\frac43\right)^n
   \left(\frac{63}{64}\right)^{\nu_1}
   (1+\varepsilon)^{-m}
 \right).
\end{align*}
Equivalently, including the suppressed overhead,
\[
 \log T_\dls
 \le
 n\log(4/3)
 -\nu_1\log(64/63)
 -m\log(1+\varepsilon)
 +o(n).
\]

The initial chain-$\dls$ cost test has failed, so
Lemma~\ref{lem:num_1_chain} gives $\nu_1>\alpha n$.
Together with $m>\sigma n$, this implies
\[
 \log T_\dls
 \le
 n\left[
   \log(4/3)
   -\alpha\log(64/63)
   -\sigma\log(1+\varepsilon)
 \right]+o(n).
\]
Exponentiation proves the stated bound. If the expression in
brackets is at most $\log c^*$, then
$T_\dls\le\mathcal O((c^*)^n)$.
\end{proof}

\begin{proof}[Proof of Lemma~\ref{lem:upperbound_simple}]
Choose
\[
 \sigma=\frac1{100},
 \qquad
 c^*=\frac{8299522295099}{6250000000000}
\]
Run $\br$ until its cost reaches
$\mathcal O((c^*)^n)$.
If the threshold is never reached,
the branching procedure solves the formula within this bound.

Otherwise, run $\textsf{Simple-}\pa_2$ or $\dls$ according to the definition of algorithm \ref{alg:framework_simple}.

If $|\mathcal I_{\mathrm{block}}|\le\sigma n$, the $\textsf{Simple-}\pa_2$ bound gives
\[
 T_\textsf{S-R2S}
 \le
 \mathcal O\!\left(
  2^{(\log c^*/\log3+6\sigma)n}
 \right).
\]
For the chosen constants,
\[
 2^{\log c^*/\log3+6\sigma}
 =1.2467384681\ldots<c^*,
\]
so this terminal also runs within
$\mathcal O((c^*)^n)$.

Finally, if $|\mathcal I_{\mathrm{block}}|>\sigma n$,
invoke $\dls$ on the selected blocks together with all remaining
singleton chains. 
Note that the chosen constants satisfy
\[
 \log(4/3)-\alpha\log(64/63)
           -\sigma\log(1+\varepsilon)
 <\log c^*.
\]
So lemma~\ref{lem:large-block-dls} therefore bounds this terminal by $\mathcal O((c^*)^n)$ as well.

Exactly one terminal is executed, and it solves the saved
original formula globally. Adding its cost to the branching
prefix gives total runtime $\mathcal O((c^*)^n)$.

Compared with Liu's baseline
\[
 c=3^{\log(4/3)/\log(64/21)}
   =1.3279235672158634\ldots,
\]
we have
\[
 c-c^*=2.3404718246896\ldots\times10^{-14}.
\]
\end{proof}

Therefore, the simple algorithm gives a strict improvement in the deterministic $3$-SAT upper bound. This demonstrates the power of exploring more complex structures that can be detected by the $\br$ procedure. In \S\ref{sec:impvbound}, we will go further down this route to provide an $\mathcal O(1.327751)$ deterministic algorithm for $3$-SAT.
\section{An Improved Deterministic Algorithm for 3-SAT}
\label{sec:newalg}

Following the same idea of exploiting complex structures we can generate from the $\br$ procedure, we can obtain significant improvement in the time complexity in deterministic $3$-SAT algorithms, as presented in the following theorem.

\begin{theorem}
\label{thm:main}
$3$-SAT on $n$ variables can be solved deterministically in time $\mathcal O(1.32736^n)$.
\end{theorem}

We present our algorithm in Algorithm~\ref{alg:Liuimproved}. It is the first strict improvement of Liu's deterministic $3$-SAT algorithm. This is the fastest deterministic local search $3$-SAT algorithm to date.

\begin{algorithm}[H]
\label{alg:Liuimproved}
\small
\setlength{\linewidth}{\hsize}
\caption{Algorithmic Framework}
\label{alg:framework}
\begin{algorithmic}[1]
\REQUIRE $3$-CNF $F$
\ENSURE a satisfying assignment or \textsf{Unsatisfiable}
\STATE $\textsf{BR}$ recursively reduces $F$ to a $2$-CNF and solves it, or returns a set $\mathcal I_\text{chain}$
\IF{$F$ is not solved in allowed branching steps}
    \STATE $(\mathcal I,\mathcal I_\text{weak},\pi)\leftarrow\ex(F,\mathcal I_\text{chain})$
    \STATE compute the cost of $\pa_2(F,\mathcal I,\pi)$ and $\dls(F,\mathcal I\cup\mathcal I_\text{weak})$
    \STATE solve $F$ with the cheaper one
\ENDIF
\end{algorithmic}
\end{algorithm}

Algorithm~\ref{alg:Liuimproved} is the framework of our algorithm. It follows this route: first try $\br$ to reduce $F$ to a $2$-CNF; if not solved, $\br$ returns a chain instance $\mathcal I_\text{chain}$. Then we call $\ex$ to construct blocks and ports, where ports are variables that we will enumerate their $0/1$ assignments in $\pa_2$ and we denote the set of them by $\pi$. Finally, $F$ is solved by $\pa_2$ or $\dls$ with $\mathcal I$.

Algorithm~\ref{alg:Liuimproved} uses the following subroutines:

\begin{algorithm}[H]
\small
\setlength{\linewidth}{\hsize}
\caption{Expansion}
\label{alg:expansion}
\begin{algorithmic}[1]
\REQUIRE $3$-CNF $F$, chain instance $\mathcal I_{\text{chain}}$
\ENSURE instances $\mathcal I$, $\mathcal I_\text{weak}$ and a port set $\pi$
        
\STATE $\mathcal I\leftarrow\{\mathcal S\in\mathcal I_{\text{chain}}|\mathcal S\text{ is constructed with truncation}\}$
\STATE $\mathcal I_\text{weak}\leftarrow\mathcal I_\text{chain}\setminus\mathcal I$
\STATE $\pi\leftarrow\varnothing$

\FOR{every $\mathcal S=\langle C_1,C_2,...,C_{|\mathcal S|}\rangle\in\mathcal I_{\text{weak}}$ with $|\mathcal S|\ne1$}
    \FOR{$i=1,...,|\mathcal S|-1$}
        \STATE $\pi\leftarrow\pi\cup\{\text{an arbitrary variable shared by $C_i$ and $C_{i+1}$}\}$
    \ENDFOR
\ENDFOR
\FOR{$i=1,2$}
    \STATE $\mathcal R\leftarrow\varnothing$
    \FOR{every $C\in F_3$ and $C$ touches $3$ distinct chains $\mathcal S\in\mathcal I_\text{weak}$}
        \IF{$C$ fails in stage-$i$ risk test}
            \STATE $\mathcal R\leftarrow\mathcal R\cup\text{the}$ output request $(C,\mathcal S,x)$ of the test.
        \ENDIF
    \ENDFOR

    \WHILE{$\mathcal R$ contains $(C_1,\mathcal S,x_1)$ and
           $(C_2,\mathcal S,x_2)$ with $x_1\ne x_2$}
        \STATE Choose such a pair in $\mathcal R$.
        \STATE $\mathcal B\leftarrow$ the block constructed from $C_1$, $C_2$ and all their touched chains $\mathcal S\in\mathcal I_\text{weak}$
        \STATE $\mathcal I\leftarrow
               \mathcal I\cup\{\mathcal B\}$.
        \STATE Remove from $\mathcal I_\text{weak}$ all the chains in $\mathcal B$ and their ports in $\pi$
        \STATE Remove from $\mathcal R$ every request whose clause
               touches $\mathcal B$.
    \ENDWHILE

    \FOR{every $(C,\mathcal S,x)\in\mathcal R$}
        \STATE Remove from $\mathcal R$ all requests that contains $\mathcal S$
        \STATE $\pi\leftarrow\pi\cup\{x\}$.
    \ENDFOR
\ENDFOR

\FOR{every $\mathcal S\in\mathcal I_{\text{weak}}$ with $|\mathcal S|=1$ and $V(\mathcal S)\cap\pi=\varnothing$}
    \STATE $\pi\leftarrow\pi\cup
           \{\text{an arbitrary variable in }\mathcal S\}$.
\ENDFOR

\FOR{every $C\in F_3$ and $\forall\mathcal S\in\mathcal I_\text{weak},C\notin\mathcal S$}
    \IF{$C$ does not touch any $\mathcal B\in\mathcal I$ and $V(C)\cap\pi=\varnothing$}
        \STATE $\mathcal B\leftarrow$ the block constructed from $C$ and all its touched chains in $\mathcal I_\text{weak}$.
        \STATE $\mathcal I\leftarrow\mathcal I\cup\{\mathcal B\}$
        \STATE Remove from $\mathcal I_\text{weak}$ all the chains in $\mathcal B$ and their ports in $\pi$
    \ENDIF
\ENDFOR

\RETURN $(\mathcal I,\mathcal I_\text{weak},\pi)$
\end{algorithmic}
\end{algorithm}

We present the stage-$i$ risk tests in natural language below. They are used in the 1st and 2nd loops, respectively.

\paragraph{Stage-$i$ risk tests.}
Consider a $3$-clause $C$ whose three variables are respectively shared with three distinct weak chains $\mathcal S_1,\mathcal S_2,\mathcal S_3$. $C$ immediately passes the test if it touches an existing block $\mathcal B\in\mathcal I$, or there is an variable of $C$ in $\pi$. If it fails, the test returns a pair $(\mathcal S,x)$, which means that $C$ asks to use its shared variable $x\in\mathcal S$ as a port.  Let $\mathcal A(\mathcal B_C)$ be the margin of the hypothetical block formed by $C$ and its touched chains.

\paragraph{Stage 1 risk test.}

This stage considers such clauses $C$ when there is no touched chain is $1$-chain and no variable of $C$ is in $\pi$. If \(\mathcal A(\mathcal B_C)<0\), it chooses an eligible pair \((\mathcal S,x)\). The pair is eligible if \(\mathcal S\) is an \textsf{n*} chain and \(x\) occurs with the same sign in $C$ and $\mathcal S$, or if \(\mathcal S\) is an \textsf{nn*} chain.\footnote{ Enumerating all possible kinds of chains under our new truncation method in Section~\ref{subsec:combine_3} shows that only these cases would cause a block with negative margin,}

\paragraph{Stage 2 risk test.}
This stage considers such clauses $C$ when there is at least $1$ touched chain is $1$-chain and no variable of $C$ is in $\pi$. If $\mathcal A(\mathcal B_C)<0$, it chooses an arbitrary touched $1$-chain $\mathcal S$ and returns $(\mathcal S,x)$, where $x$ is its shared variable with $C$.

Then the test returns the chosen chain and variable as a request.

$\ex$ is the key to constructing blocks and ports. It first separate all chains to $2$ types: paid chains are those which are constructed with truncation, and weak chains are the remaining ones. That is because truncation only happens when a chain is long enough to constrain the search space, and we immediately let each those chains to be a block. So we only have to deal with the remaining weak chains.

For a chain $\mathcal S=\langle C_1,\dots,C_{|\mathcal S|}\rangle$ the algorithm chooses an arbitrary variable shared by each two consecutive clauses as a base port. This ensures all clauses in non-$1$-chains to be reduced to a $2$-CNF, and covers more clauses outside the chains.

The subsequent procedure handles clauses that touch three distinct chains and whose resulting block has a negative margin. This is divided into $2$ stages. In stage $1$ we handles those which do not touch $1$-chains, and stage $2$, the remaining. For each such clause, we formulate a request consisting of the clause itself, a chain it touches, and a variable shared by them. For the resulting requests that ask for the distinct variables of the same chain, we package $2$ clauses in these requests and all their touched chains as a block, and remove the related ports. After this process, every chains in the requests are asked for the same variable, and we let this variable to be an extra port of that chain.

After the process, for each $1$-chains which has no ports, the algorithm let an arbitrary variable of that chain as its port. Then for each remaining clauses whose $3$ variables are not ports, it package that clause and all its touched chains as a block.

\begin{lemma}[Port Count]
\label{lem:req_port_count}
For every $\mathcal S\in\mathcal I_\text{weak}$ output by $\ex$, let $\pi^\text{base}$ be the ports constructed by Line 4-8 and $\pi^\text{req}$ be the ports constructed by Line 9-30, then
\[
|\pi^{\rm base}_\mathcal S|=|\mathcal S|-1,\qquad|\pi^{\rm req}_\mathcal S|\le1 .
\]
\end{lemma}
\begin{proof}
$|\pi^{\rm base}_\mathcal S|=|\mathcal S|-1$ is obvious by the description of $\ex$. Within a stage, after the \textsc{while} loop every request still naming $\mathcal S$ asks for the same variable, at most $1$ extra port may be added for each chains. Stage-$1$ only add ports for non-$1$-chains and stage-$2$, $1$-chains. So there are at most $1$ extra port for each chains, i.e., $|\pi^\text{req}_\mathcal S|\le1$.
\end{proof}

Then we present the solving terminal $\pa_2$ in Algorithm~\ref{alg:reduce2sat}.

\begin{algorithm}[H]
\small
\setlength{\linewidth}{\hsize}
\caption{$\pa_2$}
\label{alg:reduce2sat}
\begin{algorithmic}[1]
\REQUIRE $3$-CNF $F$, instance $\mathcal I$, port set $\pi$
\ENSURE an satisfying assignment $\alpha$ or $\textsf{Unsatisfiable}$
\STATE $A\leftarrow A(\mathcal I)\times\{0,1\}^{|\pi|}$
\FOR{every $\alpha\in A$}
    \STATE solve $F|_\alpha^{\text{UP}}$ in polynomial time
    \IF{a satisfying assignment $\alpha$ is found}
        \RETURN $\alpha$
    \ENDIF
\ENDFOR
\RETURN $\textsf{Unsatisfiable}$
\end{algorithmic}
\end{algorithm}

We will show in Lemma~\ref{lem:reduce_correct} that applying each assignments enumerated by $\pa_2$ reduces $F$ to a $2$-CNF.

\begin{lemma}
\label{lem:reduce_correct}
$(F|_{\alpha}^\text{UP})_3=\varnothing$ for every $\alpha\in (A(\mathcal I)\times\{0,1\}^{|\pi|})$.
\end{lemma}
\begin{proof}
Let $A=A(\mathcal I)\times\{0,1\}^{|\pi|}$, and $(\mathcal I,\pi)$ be the output of $\ex$. We show that $\forall C\in F_3,V(C)\cap V(A)\neq\varnothing$.
\begin{itemize}
    \item If $\exists\mathcal B\in\mathcal I,C\in\mathcal B$. All $\alpha\in A$ satisfies all blocks in $\mathcal I$, so $C$ is also satisfied,
    \item If $\forall\mathcal B\in\mathcal I,C\notin\mathcal B$. If $C$ touches any block, then $C$ is reduced to a $2$-clause by the satisfying assignment of that block. Or it must be $\exists \mathcal S\in\mathcal I_\text{weak},C\in\mathcal S$ because Line 35-41 of $\ex$ ensures that any clause not in a weak chain necessarily belongs to a block, even if that block contains only a single clause. Line 4-12 of $\ex$ ensures that enumerating the $0/1$ assignments of all ports reduces all clauses in a weak chain to a $2$-clause.
\end{itemize}
\end{proof}

It is easy to show the correctness of Algorithm~\ref{alg:Liuimproved} with Lemma~\ref{lem:reduce_correct}.

\section{The Improved Upper Bound}
\label{sec:impvbound}

In this section we prove Theorem~\ref{thm:main}. We analyze the running time of $\br$ and the two solving terminal $\pa_2$ and $\dls$, then we combine them together to get the upper bound of our algorithm. The procedure $\br$ is followed by either $\pa_2$ or $\dls$. It suffices to show that after running $\br$ for $\mathcal O(1.32736^n)$ time, it must return some structure that contains enough information such that the minimum of the runtime of $\pa_2$ and $\dls$ is $\mathcal O(1.32736^n)$.

\subsection{Auditing the Costs of the Subroutines}

Let the logarithmic cost of $\br$ be $P$, the logarithmic cost of $\pa_2$ be $S$ and the logarithmic cost of $\dls$ be $H$. 
 \[
 \max\set{P, \min \set{S, H}} \le \mathcal O(1.32736^n).
 \]
It suffices to prove lemma \ref{lem:weightsum} below, since we can bound $\log{T}=\max\set{P, \min \set{S, H}}$ with the weighted sum of $P$, $S$ and $H$.
\begin{lemma}
\label{lem:weightsum}
There exists positive weights $\mu$, $\alpha$ and $W$ such that $\mu + \alpha + W = 1$ and
    \[\mu P+\alpha S+WH\le n\log(1.32736).\]
\end{lemma}

\subsubsection{The cost of BR}

From \cite{liu2018ksat}, we know that each $n$, $p$, $tn$ and terminal $t$ incurs a branching cost of $3$, $2$, $7$ and $7/3$ respectively.
Thus the branching cost of a chain $\mathcal S$ is
\[
b(\mathcal S) = 2^{N_{\mathtt p}(\mathcal S)} 3^{1 + N_{\mathtt n}(\mathcal S)} (7 / 3)^{N_{\mathtt t}(\mathcal S)}
\]
where $N_s(\mathcal S)$ is the number of $s \in \set{n, p, t}$ in $\zeta(\mathcal S)$. If $\mathcal S$ is constructed with truncation, we multiply $b(\mathcal S)$ with an extra factor $2$, inserted exactly when the paid reduction action actually fires.

For a block $\mathcal B$, its branching cost is the product of the branching costs of all the chains it contains, i.e., $b(\mathcal B)=\prod_{\mathcal S\in\mathcal B}b(\mathcal S)$.

Then, we obtain the total cost of $\br$ as the following lemma.

\begin{lemma}[BR cost]
\label{lem:br_time}
Let $T_\br$ be the running time of $\br$, we have $T_\br\le O^*(e^P)$, where
\[
P=\sum_{\mathcal S\in\mathcal I_\text{weak}}\log b(\mathcal S)+\sum_{\mathcal B\in\mathcal I}\log b(\mathcal B).
\]
\end{lemma}
\begin{proof}
By \cite{liu2018ksat} we have $T_\br\le\prod_{\mathcal S\in\mathcal I_\text{chain}}b(\mathcal S)$. By the definition of Algorithm~\ref{alg:expansion} we have $\mathcal I_\text{chain}=\mathcal I_\text{weak}\cup\mathcal S(\mathcal I)$, where $\mathcal S(\mathcal I)$ is the set of chains in all blocks in $\mathcal I$. Then the conclusion follows.
\end{proof}

\subsubsection{The cost of PA2}

\begin{lemma}[PA2 cost]
\label{lem:reduce_time}
Let $T_{\pa_2}$ be the running time of $\pa_2$, we have $T_{\pa_2}\le O^*(e^S)$, where
\[
S=\sum_{\mathcal S\in\mathcal I_\text{weak}}|\mathcal S|\log2+\sum_{\mathcal B\in\mathcal I}\log|A(\mathcal B)|.
\]
\end{lemma}

\begin{proof}
The enumeration first chooses a satisfying assignment
for each block $\mathcal B\in\mathcal I$. These choices contribute
\[
\prod_{\mathcal B\in\mathcal I} |A(\mathcal B)|.
\]

Then, $\pa_2$ has to fix the variables in $\pi$. That part contributes cost $2^{|\pi|}$. Thus the enumeration count is
\[
2^{|\pi|}\prod_{\mathcal B\in\mathcal I} |A(\mathcal B)|.
\]

Only weak chains have ports and each weak chain $\mathcal S$ has at most $|\mathcal S|$ ports, so $|\pi|\le\sum_{\mathcal S\in\mathcal I_\text{weak}}|\mathcal S|$. Take logarithm to obtain $S$.
\end{proof}

\subsubsection{The cost of DLS}

\begin{definition}[Weight]
\label{def:weight}
Given a component $\mathcal K$, define its \emph{weight} $Q(\mathcal K)$ by
\[
Q(\mathcal K)=|V(\mathcal K)|\log{4/3}+\log{\lambda_\mathcal K},
\]
\end{definition}

The weight of a component measures the amount of information it contains. The larger the weight, the more information it can provide, and the smaller the covering code. Then, we can rewrite the logarithm running time of $\dls$ as the following lemma.

\begin{lemma}
\label{lem:dls_time}
Let $T_\dls$ be the running time of $\dls$, we have $T_\dls\le O^*(e^H)$, where
\[
H=n\log\frac43-\sum_{\mathcal S\in\mathcal I_\text{weak}}Q(\mathcal S)-\sum_{\mathcal B\in\mathcal I}Q(\mathcal B).
\]
\end{lemma}

\subsection{Combination of the three algorithm}
\label{subsec:combine_3}

\begin{lemma}
The whole algorithm runs in time $T\le\mathcal O(e^{\mathcal En})$, where
\[
\mathcal E= \frac{\max\{P,\min\{S,H\}\}}{n}.
\]
\end{lemma}

\begin{proof}
By lemma~\ref{lem:br_time}, lemma~\ref{lem:reduce_time} and lemma~\ref{lem:dls_time} the conclusion follows.
\end{proof}

We first define the following quantity to measure the slackness that a component can provide for the running time.

\begin{definition}
For a weak chain $\mathcal S$, define its \emph{margin} as
\[
\mathcal A(\mathcal S)=WQ(\mathcal S)-\alpha|\mathcal S|\log2-\mu\log b(\mathcal S),
\]
and for a block $\mathcal B$, define its \emph{margin} as
\[
\mathcal A(\mathcal B)=WQ(\mathcal B)-\alpha\log{|A(\mathcal B)|}-\mu\log b(\mathcal B).
\]
\end{definition}

In the definition of margin, the terms corresponding to $W,\alpha$ and $\mu$ respectively represent the effects of a block or a chain on the values of $H$, $S$, and $P$. This is why it can be used to quantity the contribution of a chain or block to the whole running time.

\begin{lemma}
\label{lem:margin_non_negative}
There exist $W,\alpha,\mu$ such that for every block $\mathcal B\in\mathcal I$, $\mathcal A(\mathcal B)\ge0$ and for every weak chain $\mathcal S\in\mathcal I_\text{weak}$, $\mathcal A(\mathcal S)\ge0$.
\end{lemma}
\begin{proof}
We explicitly provide such values of $W,\alpha$ and $\mu$, and prove that $\mathcal A(\mathcal S)\ge0$ holds under these values.

Let $\mathcal A(\textsf{*})=0$. This gives
\[
W\log\frac{64}{63}-\alpha\log2-\mu\log3=0.
\]

Choose $\frac{\alpha}{W}=\frac{103}{25000}$ gives
\[
W=\frac{\log3}{\log\frac{64}{21}+\frac{103}{25000}\log\frac{3}{2}},\quad
\alpha=\frac{\frac{103}{25000}\log3}{\log\frac{64}{21}+\frac{103}{25000}\log\frac{3}{2}},\quad
\mu=\frac{\log\frac{64}{63}-\frac{103}{25000}\log2}{\log\frac{64}{21}+\frac{103}{25000}\log\frac{3}{2}},
\]
with numerical values $W=0.9843921$, $\alpha=0.0040557$ and $\mu=0.0115522$.

We first prove the chain inequality by examining the change in margin
when an admissible extension is appended. Since the proof of the inequality $\mathcal A(\mathcal B)\ge0$ is long, we prove it in Appendix~\ref{app:block-contributions}. 

Define
\[
\widehat Q(\mathcal S)
=v(\mathcal S)\log\frac43-\log M(\mathcal S),
\qquad
\widehat{\mathcal A}(\mathcal S)
=W\widehat Q(\mathcal S)
-\alpha\,|\mathcal S|\log 2-\mu\log b(\mathcal S).
\]
For a chain $\mathcal S$ and an admissible extension
$w\in\{\textsf{n},\textsf{p},\textsf{t}\}$, write
$m_w=M(\mathcal S+w)/M(\mathcal S)$. Let $d_w$ be the number of added
variables, $\beta_w$ the branching-cost multiplier. Subtracting the two
margins gives the exact identity
\[
\Delta_w
=W\left(d_w\log\frac43-\log m_w\right)
  -\alpha\log 2-\mu\log\beta_w .
\]
For $w=\textsf{n},\textsf{p},\textsf{t}$, $(d_w,\beta_w)=(2,3),(2,2),(1,7/3)$.
Thus upper bounds on $m_w$ give lower bounds on $\Delta_w$.
Theorem~\ref{thm:one-clause-recurrence}, together with
Lemmas~\ref{lem:ratio-transition} and~\ref{lem:ratio-bounds}, supplies
the bounds on $m_w$. Positive increments, together with the separately
checked initial chains, establish $\widehat{\mathcal A}\ge0$.
Additionally, $Q\ge\widehat Q$ holds by
Corollary~\ref{cor:Qlower_bound}, hence
$\mathcal A\ge\widehat{\mathcal A}\ge0$.

For every later extension, Theorem~\ref{thm:one-clause-recurrence} and
Lemma~\ref{lem:ratio-transition} give
\[
m_\textsf{n}\le\frac{173}{99},\qquad
m_\textsf{p}\le\frac{79}{45},\qquad
m_\textsf{t}\le\frac{25}{19}.
\]
The corresponding gains in $Q$ for $\textsf{n}$, $\textsf{p}$ and $\textsf{t}$ are at least
\[
\frac{176}{173},\qquad
\frac{80}{79},\qquad
\frac{76}{75}.
\]
The margin increments therefore satisfy
\[
\begin{aligned}
\Delta_{\textsf n}
&\ge W\log\frac{176}{173}-\alpha\log2-\mu\log3>0,\\
\Delta_{\textsf p}
&\ge W\log\frac{80}{79}-\alpha\log2-\mu\log2>0,\\
\Delta_{\textsf t}
&\ge W\log\frac{76}{75}-\alpha\log2-\mu\log\frac73>0.
\end{aligned}
\]
whose values are $0.00142$, $0.00157$, $0.00044$ and $0.00467$ respectively.

Our chosen $W,\alpha$ and $\mu$ satisfies $\mathcal A(\textsf{*})=0$. Every longer completed chain is obtained from \textsf{*} by repeatedly appending \textsf{n}, \textsf{p}, or \textsf{t}. Each such extension has strictly positive margin increment.
Therefore, by induction,
\[
\mathcal A(w^*)\ge0
\]
for every completed chain, with equality exactly for $\textsf{*}$.
\end{proof}

This gives an intuition that in the worst case, $\mathcal I_\text{chain}$ should contain only \textsf{*} because it is the only chain with $\mathcal A(\mathcal S)=0$.

Then we can truncate a chain in $\br$ when $\mathcal A(\mathcal S)\ge\mu\log2$ if treat the chain as a block that is consists of only $1$ chain and use the margin of a block for it, without break Lemma ~\ref{lem:margin_non_negative}. Enumerating all possible kinds of chains shows that the number of possible kinds of chains and blocks is finite and thus they can be hardwired.

We give the upper bound of $\mathcal E$ as the following lemma.

\begin{lemma}
\label{lem:upperbound}
Let $\br$ stop at the first permitted checkpoint satisfying
\[
\Phi(E_t):
\frac{\sum_{\mathcal S\in\mathcal I_\text{chain}}\log b(\mathcal S)}{n}
\ge E_t,
\]
then
\[
\mathcal E\le E_t+o(1),
\]
where
\[
E_t=
\frac{\log 3\cdot\log\frac43}
{\log{\frac{64}{21}}+\frac{103}{25000}\log\frac{3}{2}}.
\]
\end{lemma}

\begin{proof}
Since the blocks and chains have disjoint supports, the three cost expressions combine term by term as
\[
\begin{aligned}
\mu P+\alpha S+WH
&=Wn\log\frac43
+\sum_{\mathcal S\in\mathcal I_\text{weak}}\Bigl(\mu\log b(\mathcal S)+\alpha|\mathcal S|\log2-WQ(\mathcal S)\Bigr)\\
&\qquad+\sum_{\mathcal B\in\mathcal I}\Bigl(\mu\log b(\mathcal B)+\alpha\log|A(\mathcal B)|-WQ(\mathcal B)\Bigr)\\
&=Wn\log\frac43-\sum_{\mathcal S\in\mathcal I_\text{weak}}\mathcal A(\mathcal S)-\sum_{\mathcal B\in\mathcal I}\mathcal A(\mathcal B).
\end{aligned}
\]

By Lemma~\ref{lem:margin_non_negative} we have
$\mathcal A(\mathcal K)\ge0$ for every component. Therefore we have the desired upper bound
\[
\mu P+\alpha S+WH\le Wn\log\frac43.
\]

Consequently,
\[
W\log\frac43=E_t,
\]
and the weighted inequality becomes
\[
\mu P+\alpha S+WH\le E_tn.
\]

At the stopping checkpoint, $P\ge E_tn$.
It follows that
\[
\begin{aligned}
\alpha S+WH
&\le E_tn-\mu P\\
&\le(1-\mu)E_tn\\
&=(\alpha+W)E_tn.
\end{aligned}
\]
Since $\alpha,W>0$,
\[
\min\{S,H\}
\le\frac{\alpha S+WH}{\alpha+W}
\le E_tn.
\]
Thus at least one of the two terminal procedures has
the required logarithmic cost.

We next bound the branching work. Immediately before the branching operation that triggers termination, the stopping condition is false. Each operation between consecutive stopping tests multiplies the charged branch product by at most $7$, and the paid action contributes exactly one factor $2$. Consequently,
\[
E_tn\le P<E_tn+\log7.
\]

Since every internal node of the branching prefix has at least two children, the number of its leaves is at most the largest root-to-leaf branch product, and each factor of a root-to-leaf path occurs at most once in $e^P$, so that product is dominated by $e^{P}$; every abandoned sibling subtree is charged to the truncation factor of the chain that generated it. By Lemma~\ref{lem:br_time}, this bound on the charged
branch products gives a bound
$\mathcal O(e^{E_tn})$ on the complete branching prefix,
including all previously explored branches.

After $\br$ stops, alternate computation steps of $\pa_2$
and $\dls$. This takes at most twice the running time of
the faster procedure, up to constant overhead. Since
both procedures decide the original formula, the first
completed procedure terminates the whole algorithm.
The terminal running time is therefore
\[
\mathcal O\!\left(
 \exp\bigl(\min\{S, H\}\bigr)
\right)
\le \mathcal O(e^{E_tn}).
\]

Adding the branching and terminal costs proves the complexity
\[
O(e^{E_tn}).
\]
Finally,
\[
\mathcal E
=\frac{\max\{P,\min\{S,H\}\}}{n}
\le E_t+O(1/n)
=E_t+o(1).
\]
\end{proof}

We finally obtain our main result on $3$-SAT as Theorem~\ref{thm:best_running_time}.

\begin{theorem}
\label{thm:best_running_time}
There exists a deterministic algorithm for $3$-SAT that runs in time $\mathcal O(e^{E_tn})$, with $e^{E_t}=1.3273599467\ldots$
\end{theorem}
\section{Generalize to \texorpdfstring{$k\ge 4$}{kge4}}

\subsection{Overview}

Here we generalize our $3$-SAT algorithm to the case where $k
\ge 4$, by exploiting the natural idea of reducing $k$-SAT to $(k-1)$-SAT by a partial assignment of a maximum instance of $1$-chains or blocks. Specifically, our algorithm combines three ways of solving the same formula. Its runtime is presented in Theorem~\ref{thm:general-k-sat}. Both $\br_k$ and $\pa(k-1)$ enumerate partial assignments and solve the remaining $(k-1)$-CNF recursively. Their costs therefore depend on $c_{k-1}$. $\dls$ searches the original formula using a covering code, so its cost depends on $q_k$ and the characteristic values of the components. The recurrence brings these bounds together, with the improved $3$-SAT base $c_3$ as the base case.

\begin{theorem}[Theorem~\ref{thm:general-k-sat}, informal]
We can generalize our framework to obtain a general $k$-SAT algorithm, whose runtime base $c_k$ can be written in a recurrence of $c_{k - 1}$.

For example, evaluating the recurrence gives
\[
c_4\approx c_4^{\mathrm{Liu}} - 3.3\times10^{-4},\quad
c_5\approx c_5^{\mathrm{Liu}} - \approx1.0\times10^{-4},\quad
c_6  \approx c_6^{\mathrm{Liu}}-3.4\times10^{-5}.
\]
Above, $c_k$ is the runtime base of our algorithm solving $k$-SAT, and $c_k^{\mathrm{Liu}}$ is the runtime base of Liu's algorithm solving $k$-SAT~\cite{liu2018ksat}.
\end{theorem}

\begin{algorithm}[H]
\small
\caption{Deterministic $k$-SAT: $\textsf{Solve}_k$, $k\ge4$}
\label{alg:ksat-framework}
\begin{algorithmic}[1]
\REQUIRE A normalized $k$-CNF $F$ with $k\ge 4$
\ENSURE A satisfying assignment of $F$ or \textsf{Unsatisfiable}

\IF{$F$ contains an empty clause}
    \RETURN \textsf{Unsatisfiable}
\ENDIF
\IF{$F$ contains no clause of size $k$}
    \RETURN $\textsf{Solve}_{k-1}(F)$
\ENDIF

\STATE Greedily construct a maximal instance $\mathcal I_1$
       of variable-disjoint $1$-chains with clauses of $F$
\STATE $\mathcal I_{\mathrm{block}}
       \leftarrow\textsf{ConstructBlock}_k(F,\mathcal I_1)$
\STATE $\mathcal U\leftarrow
       \{\mathcal S\in\mathcal I_1:
       V(\mathcal S)\cap V(\mathcal I_{\mathrm{block}})
       =\varnothing\}$
\STATE $\mathcal I\leftarrow
       \mathcal U\cup\mathcal I_{\mathrm{block}}$
\STATE Compute the costs $P$, $S$ and $H$

\IF{$P\le\min\{S,H\}$}
    \RETURN $\br(F, \mathcal I_1)$
\ELSIF{$S\le H$}
    \RETURN $\mathsf{PA(k - 1)}$ $(F, \mathcal I_{\mathrm{block}}, \mathcal U)$
\ELSE
    \RETURN $\textsf{DLS}(F,\mathcal I)$
\ENDIF
\end{algorithmic}
\end{algorithm}

Algorithm~\ref{alg:ksat-framework} extends our $1.32736^n$ algorithm for $3$-SAT by reducing the clause width recursively, with $\textsf{Solve}_3$ as the base case. It first constructs a maximal collection of independent $1$-chains and groups some of them into blocks. These components determine running-time bounds for $\br$, $\textsf{PA}_{(k-1)}$, and $\dls$. $\br_k$ and $\pa_{(k-1)}$ produce $(k-1)$-SAT instances for the recursive solver, while $\dls$ searches the original formula using a covering code built from the components. The main difference from the $3$-SAT framework is when this choice is made. For $3$-SAT, $\br$ constructs chains during branching, and whichever runs faster of $\pa_2$ or $\dls$ is used if $\br$ does not decide the formula. Here, the components are constructed in polynomial time for fixed $k$, so the algorithm compares all three bounds before starting the exponential computation.


\begin{algorithm}[H]
\small
\caption{$\textsf{ConstructBlock}_k$}
\label{alg:construct-block-k}
\begin{algorithmic}[1]
\REQUIRE A normalized $k$-CNF $F$ and a maximal instance
         $\mathcal I_1$ of independent $1$-chains
\ENSURE An instance $\mathcal I_{\mathrm{block}}$
        of variable-disjoint blocks

\STATE $\mathcal I_{\mathrm{block}}\leftarrow\varnothing$
\FOR{each $C\in F$ with $|C|=k$
     and $\langle C\rangle\notin\mathcal I_1$}
    \STATE $\mathcal B\leftarrow
       \{C\}\cup
       \{C_0:
          \langle C_0\rangle\in\mathcal I_1
          \text{ and }V(C_0)\cap V(C)\ne\varnothing\}$
    \IF{$V(\mathcal B)\cap
         V(\mathcal I_{\mathrm{block}})=\varnothing$}
        \STATE $\mathcal I_{\mathrm{block}}\leftarrow
           \mathcal I_{\mathrm{block}}\cup\{\mathcal B\}$
    \ENDIF
\ENDFOR
\RETURN $\mathcal I_{\mathrm{block}}$
\end{algorithmic}
\end{algorithm}

Algorithm~\ref{alg:construct-block-k} extends the block construction used in the $3$-SAT algorithm. For each $k$-clause outside $\mathcal I_1$, it forms a block consisting of that clause and every $1$-chain sharing one of its variables. The block is selected if its variables are disjoint from those of previously selected blocks. Each touched $1$-chain is included in full. Together with the maximality of the selected collection, this ensures that every $k$-clause outside $\mathcal I_1$ contains a variable of a selected block. Assigning the block variables therefore shortens these clauses for $\pa_{(k-1)}$. The blocks also help $\dls$ by restricting the assignments that a satisfying assignment of the formula can use. Since a central clause touches at most $k$ independent $1$-chains, each block contains at most $k^2$ variables.

\begin{algorithm}[H]
\small
\caption{$\textsf{\br}_k$}
\label{alg:enumerate-solve-k}
\begin{algorithmic}[1]
\REQUIRE A formula $F$ and a maximal singleton chain instance $\mathcal I_1$
\ENSURE A satisfying assignment of $F$,
        or \textsf{Unsatisfiable} if no such extension exists
        
\STATE $A\leftarrow
       \displaystyle\prod_{\mathcal S\in\mathcal I_1}
       A(\mathcal S)$
\FOR{each $\alpha\in A$}
    \STATE $\beta\leftarrow\textsf{Solve}_{k-1}(F|_\alpha)$
    \IF{$\beta\ne\textsf{Unsatisfiable}$}
        \RETURN $\alpha\cup\beta$
                
    \ENDIF
\ENDFOR
\RETURN \textsf{Unsatisfiable}
\end{algorithmic}
\end{algorithm}

Algorithm~\ref{alg:enumerate-solve-k} enumerates the $2^k-1$ satisfying assignments of each $1$-chain in $\mathcal I_1$, assigning all variables in these chains. Maximality of $\mathcal I_1$ means that every other $k$-clause shares a variable with one of them. Each $k$-clause is therefore satisfied or shortened, leaving a $(k-1)$-CNF for the recursive solver. The underlying idea is the same as in the earlier branching algorithm: assign variables to reduce the clause width. The polynomial-time $2$-SAT solver is now replaced by $\textsf{Solve}_{k-1}$. This version of $\br$ works with the $1$-chains instance and completes the enumeration. It does not grow longer chains.

\begin{algorithm}[H]
\small
\setlength{\linewidth}{\hsize}
\caption{Partial assignments to $(k-1)$-CNF:
         $\textsf{PA}{(k-1)}$}
\label{alg:pa-kminus1}
\begin{algorithmic}[1]
\REQUIRE $k$-CNF $F$, blocks
         $\mathcal I_{\mathrm{block}}$ returned by
         $\textsf{ConstructBlock}_k$, and $\mathcal U$, the singleton chains uncovered by $\mathcal I_{\mathrm{block}}$
\ENSURE A satisfying assignment of $F$
        or \textsf{Unsatisfiable}

\STATE Choose one variable $x_{\mathcal S}\in V(\mathcal S)$
       for each $\mathcal S\in\mathcal U$
\STATE $V_P\leftarrow
       \{x_{\mathcal S}:\mathcal S\in\mathcal U\}$
\STATE $A\leftarrow
       A(\mathcal I_{\mathrm{block}})
       \times\{0,1\}^{V_P}$

\FOR{every $\alpha\in A$}
    \STATE Compute $G=\mathrm{UP}(F|_\alpha)$,
           recording extra assignments made in this process, $\tau$
    \IF{$\perp \in G$}
        \STATE \textbf{continue}
    \ENDIF
    \STATE $\beta\leftarrow\textsf{Solve}_{k-1}(G)$
    \IF{$\beta\ne\textsf{Unsatisfiable}$}
        \RETURN $\alpha\cup\tau\cup\beta$
    \ENDIF
\ENDFOR
\RETURN \textsf{Unsatisfiable}
\end{algorithmic}
\end{algorithm}

$\pa_{(k-1)}$ generalizes $\pa2$ by assigning every variable in the selected blocks and only one variable in each uncovered $1$-chain. The selected blocks ensure that every $k$-clause outside $\mathcal I_1$ is satisfied or shortened. The chosen variable does the same for each uncovered $1$-chain. Thus the residual formula is a $(k-1)$-SAT formula, even before unit propagation. The main change is the final solving step: the residual formula now requires $\textsf{Solve}_{k-1}$. The cost bound must therefore balance fewer choices on uncovered chains against more variables left for the recursive call.

\subsection{Counting the Costs}

We keep $P$, $S$, and $H$ as total logarithmic costs.
Write
\[
s=|\mathcal I_1|,
\qquad
h=|\mathcal U|.
\]
Reserve type $1$ for a single $k$-clause, and let
$\mathcal J_k$ index the block types.
For each $i\in\mathcal J_k$, let $\nu_i$ be the number of
selected blocks of type $i$, and let $A_i$ and $\lambda_i$
be its solution space and characteristic value.
Write $r_i$ for the number of $1$-chains in the block and
$f_i$ for the number of its variables outside
$V(\mathcal I_1)$. Its number of variables is
\[
\eta_i=kr_i+f_i.
\]
Since the selected blocks and uncovered $1$-chains are
variable-disjoint,
\[
s=h+\sum_{i\in\mathcal J_k}r_i\nu_i,
\qquad
n'=n-kh-\sum_{i\in\mathcal J_k}\eta_i\nu_i,
\]
where $n'$ counts variables outside
$\mathcal I=\mathcal U\cup\mathcal I_{\mathrm{block}}$.

\subsubsection{The Cost of \texorpdfstring{$\br_k$}{brk}}

Each $1$-chain has $2^k-1$ satisfying assignments.
Thus $\br_k$ enumerates $(2^k-1)^s$ partial assignments,
each fixing all $ks$ variables of $\mathcal I_1$.
Maximality of $\mathcal I_1$ ensures that every remaining
$k$-clause is satisfied or shortened. Each residual
formula is therefore a $(k-1)$-CNF on at most $n-ks$
variables, giving
\[
T_{\br_k}
=\mathcal O\!\left((2^k-1)^s c^{\,n-ks}\right).
\]
We thus have the lemma:
\begin{lemma}
The cost of $\br_k$ is
    \begin{equation}
\label{eq:br-k-cost}
P=s\log(2^k-1)+(n-ks)\log c.
\end{equation}
This cost includes all recursive calls to
$\textsf{Solve}_{k-1}$.
\end{lemma}

\subsubsection{The Cost of PA(k-1)}

The procedure enumerates a satisfying assignment of each
selected block and both values of one chosen variable
in each uncovered $1$-chain. The number of choices is
\[
2^h\prod_{i\in\mathcal J_k}|A_i|^{\nu_i}.
\]
Each choice fixes
$h+\sum_{i\in\mathcal J_k}\eta_i\nu_i$ variables.
The block construction ensures that every $k$-clause
outside $\mathcal I_1$ contains a selected block variable,
while each uncovered $1$-chain contains its chosen variable.
The residual formula therefore has width at most $k-1$
and at most
\[
n-h-\sum_{i\in\mathcal J_k}\eta_i\nu_i
=n'+(k-1)h
\]
variables. Unit propagation can only reduce this number.
Hence
\[
T_{\textsf{PA}(k-1)}
=\mathcal O\!\left(
2^h\prod_{i\in\mathcal J_k}|A_i|^{\nu_i}
c^{\,n'+(k-1)h}
\right),
\]
and we have
\begin{lemma}
The cost of $\pa(k-1)$ is
    \begin{equation}
    \label{eq:pa-k-cost}
    S=h\log2
    +\sum_{i\in\mathcal J_k}\nu_i\log|A_i|
    +\bigl(n'+(k-1)h\bigr)\log c.
    \end{equation}
\end{lemma}

The last term accounts for solving the residual instances.
For $\pa2$, this term vanishes because $2$-SAT is solvable
in polynomial time.

\subsubsection{The Cost of DLS}

DLS uses the instance
$\mathcal I=\mathcal U\cup\mathcal I_{\mathrm{block}}$.
Put
\[
q_k=\frac{2(k-1)}{k}.
\]
Liu's \cite{liu2018ksat} characteristic value for a $1$-chain formed by a
$k$-clause is
\[
\lambda_1
=\frac{k^k}{(2k-2)^k-(k-2)^k}.
\]
Applying Liu's covering-code bound
\cite{liu2018ksat} to the uncovered $1$-chains and selected
blocks gives
\[
T_{\dls}
=\mathcal O\!\left(
q_k^{\,n'}\lambda_1^{-h}
\prod_{i\in\mathcal J_k}\lambda_i^{-\nu_i}
\right).
\]
Here each $\lambda_i$ is supported by a feasible
characteristic distribution on $A_i$.
Taking logarithms yields the lemma
\begin{lemma}
The cost of $\dls$ is
    \begin{equation}
    \label{eq:dls-k-cost}
    H=n'\log q_k-h\log\lambda_1
    -\sum_{i\in\mathcal J_k}\nu_i\log\lambda_i.
    \end{equation}
\end{lemma}

Equivalently, extend the component weight from the
$3$-SAT analysis by setting
\[
Q(\mathcal K)
=|V(\mathcal K)|\log q_k+\log\lambda_{\mathcal K}.
\]
If $\mathcal S_1$ denotes the singleton type and
$\mathcal B_i$ denotes block type $i$, then
\[
H=n\log q_k-hQ(\mathcal S_1)
-\sum_{i\in\mathcal J_k}\nu_iQ(\mathcal B_i).
\]
Thus each component reduces the DLS logarithmic cost by
its weight, exactly as in the $3$-SAT analysis.

\subsection{The Upper Bound on Runtime}

The logarithm runtime of Algorithm~\ref{alg:construct-block-k} is upper bounded by $\min\set{P, S, H}$. We use the same weighted sum idea.
\begin{lemma}
\label{lem:weighted-k-bound}
    Let
    \[
    \mu,\alpha,W\ge0,\qquad \mu+\alpha+W=1.
    \]
    If \(\gamma\) satisfies
    \[
    c^{\mu+\alpha}q_k^W\le\gamma,
    \]
    \[
    |A_1|^\mu(2c^{k-1})^\alpha\lambda_1^{-W}
    \le\gamma^k,
    \]
    and, for every \(i\in\mathcal J_k\),
    \[
    |A_1|^{\mu r_i}c^{\mu f_i}
    |A_i|^\alpha\lambda_i^{-W}
    \le\gamma^{\eta_i},
    \]
    then
    \[
    \min\{P,S,H\}\le n\log\gamma.
    \]
\end{lemma}

\begin{proof}

Substituting the cost expressions
\eqref{eq:br-k-cost}--\eqref{eq:dls-k-cost}
and collecting the contributions of outside variables,
uncovered $1$-chains, and blocks yields
\[
\begin{aligned}
\mu P+\alpha S+WH
&={}&n'\log\!\left(c^{\mu+\alpha}q_k^W\right)\\
&+h\log\!\left(
|A_1|^\mu(2c^{k-1})^\alpha\lambda_1^{-W}
\right)\\
&+\sum_{i\in\mathcal J_k}\nu_i
\log\!\left(
|A_1|^{\mu r_i}c^{\mu f_i}
|A_i|^\alpha\lambda_i^{-W}
\right).
\end{aligned}
\]
By the hypotheses, the three kinds of contributions are
at most $n'\log\gamma$, $kh\log\gamma$, and
$\eta_i\nu_i\log\gamma$, respectively. Hence
\[
\begin{aligned}
\min\{P,S,H\}
&\le\mu P+\alpha S+WH\\
&\le
\left(n'+kh+\sum_{i\in\mathcal J_k}\eta_i\nu_i\right)
\log\gamma\\
&=n\log\gamma.
\end{aligned}
\]
\end{proof}

Now we are ready to prove the main theorem of this section.

\begin{theorem}[Deterministic algorithm for $k$-SAT]
\label{thm:general-k-sat}
Let $c_3=1.3273599467376625114\ldots$.
For every fixed $k\ge4$, there exists
a deterministic algorithm solving $k$-SAT in time
\[
\mathcal O\!\left((c_k)^n\right),
\]
where,
\[
c_k=
\min_{\substack{\mu,\alpha,W\ge0\\\mu+\alpha+W=1}}
\max\left\{
\begin{aligned}
&c_{k - 1}^{\mu+\alpha}q_k^W,\\
&\left(
 |A_1|^\mu(2c_{k - 1}^{k-1})^\alpha\lambda_1^{-W}
 \right)^{1/k},\\
&\max_{i\in\mathcal J_k}
 \left(
 |A_1|^{\mu r_i}c_{k - 1}^{\mu f_i}
 |A_i|^\alpha\lambda_i^{-W}
 \right)^{1/\eta_i}
\end{aligned}
\right\}.
\]
\end{theorem}

\begin{proof}
    We proceed by induction on $k$.
The case $k=3$ follows from
Theorem~\ref{thm:best_running_time}, and
$c_3<4/3=q_3$.

Fix $k\ge4$ and assume the claim holds for $k-1$.
Set $c=c_{k-1}$. Then
\[
c\le q_{k-1}<q_k.
\]

We first record a bound on characteristic values.
For any component on $\eta$ variables with characteristic
distribution $\pi$ and value $\lambda$,
\[
\lambda\ge q_k^{-\eta}.
\]
To see this, extend $\pi$ by zero to $\{0,1\}^{\eta}$,
let $u$ be the uniform distribution on this cube, and put
\[
K(a,b)=\left(\frac1{k-1}\right)^{d(a,b)}.
\]
The matrix $K$ is positive definite, since it is the
$\eta$-fold tensor product of
\[
\begin{pmatrix}
1 & 1/(k-1)\\
1/(k-1) & 1
\end{pmatrix}.
\]
Moreover, $Ku=q_k^{-\eta}\mathbf1$.
The characteristic equations and
$\mathbf1^{\mathsf T}(\pi-u)=0$ therefore imply
\[
\lambda
=\pi^{\mathsf T}K\pi
=q_k^{-\eta}
+(\pi-u)^{\mathsf T}K(\pi-u)
\ge q_k^{-\eta}.
\]

In particular,
\[
\lambda_1\ge q_k^{-k},
\qquad
\lambda_i\ge q_k^{-\eta_i}
\quad(i\in\mathcal J_k).
\]
Choosing $(\mu,\alpha,W)=(0,0,1)$ in the recurrence gives
$c_k\le q_k$.
Conversely, for every admissible choice of weights,
\[
c^{\mu+\alpha}q_k^W
=c^{1-W}q_k^W\ge c,
\]
so $c_k\ge c$.

An empty clause is detected in polynomial time.
If $F$ contains no $k$-clause, the algorithm calls
$\textsf{Solve}_{k-1}$ directly. By induction and
$c\le c_k$, this takes time $\mathcal O(c_k^n)$.
We may therefore assume that the component construction
is performed.

For fixed $k$, the set $\mathcal J_k$ is finite.
The objective in the recurrence is continuous on the
compact set
\[
\{(\mu,\alpha,W)\ge0:\mu+\alpha+W=1\}.
\]
Thus its minimum is attained.
Choose minimizing weights. By the definition of $c_k$,
these weights satisfy all three hypotheses of
Lemma~\ref{lem:weighted-k-bound} with $\gamma=c_k$.
Consequently,
\[
\min\{P,S,H\}\le n\log c_k.
\]
The runtime of our algorithm is bounded by $\mathcal O(\exp(\min\{P,S,H\}))$, and thus by $\mathcal O(c_k^n)$.
\end{proof}

\bibliographystyle{alpha}
\bibliography{reference}

@article{hertli20143,
  author  = {Timon Hertli},
  title   = {{3-SAT} Faster and Simpler---{Unique-SAT} Bounds for {PPSZ} Hold in General},
  journal = {SIAM Journal on Computing},
  volume  = {43},
  number  = {2},
  pages   = {718--729},
  year    = {2014},
  doi     = {10.1137/120868177}
}

@incollection{hofmeister2002probabilistic,
  title={A Probabilistic 3-SAT Algorithm Further Improved},
  author={Hofmeister, Thomas and Sch{\"o}ning, Uwe and Schuler, Rainer and Watanabe, Osamu},
  booktitle={19th Annual Symposium on Theoretical Aspects of Computer Science, STACS 2002},
  pages={192--202},
  year={2002},
  publisher={Springer}
}

@article{DBLP:journals/jacm/PaturiPSZ05,
  author  = {Ramamohan Paturi and Pavel Pudl{\'a}k and Michael E. Saks and Francis Zane},
  title   = {An Improved Exponential-Time Algorithm for {$k$-SAT}},
  journal = {Journal of the ACM},
  volume  = {52},
  number  = {3},
  pages   = {337--364},
  year    = {2005},
  doi     = {10.1145/1066100.1066101}
}

@inproceedings{schoning1999probabilistic,
  author    = {Uwe Sch{\"{o}}ning},
  title     = {A Probabilistic Algorithm for k-SAT and Constraint Satisfaction Problems},
  booktitle = {40th Annual Symposium on Foundations of Computer Science, {FOCS} 1999},
  pages     = {410--414},
  year      = {1999}
}

@article{dantsin2002deterministic,
  author  = {Evgeny Dantsin and Andreas Goerdt and Edward A. Hirsch and Ravi Kannan and Jon Kleinberg and Christos Papadimitriou and Prabhakar Raghavan and Uwe Sch{\"o}ning},
  title   = {A Deterministic {$(2-2/(k+1))^n$} Algorithm for {$k$-SAT} Based on Local Search},
  journal = {Theoretical Computer Science},
  volume  = {289},
  number  = {1},
  pages   = {69--83},
  year    = {2002},
  doi     = {10.1016/S0304-3975(01)00174-8}
}

@inproceedings{moser2011full,
  author    = {Robin A. Moser and Dominik Scheder},
  title     = {A Full Derandomization of {Sch{\"o}ning}'s {$k$-SAT} Algorithm},
  booktitle = {Proceedings of the 43rd Annual ACM Symposium on Theory of Computing},
  pages     = {245--252},
  publisher = {ACM},
  year      = {2011},
  doi       = {10.1145/1993636.1993670}
}

@article{DBLP:journals/algorithmica/MakinoTY13,
  author  = {Kazuhisa Makino and Suguru Tamaki and Masaki Yamamoto},
  title   = {Derandomizing the {HSSW} Algorithm for {3-SAT}},
  journal = {Algorithmica},
  volume  = {67},
  number  = {2},
  pages   = {112--124},
  year    = {2013},
  doi     = {10.1007/s00453-012-9741-4}
}

@inproceedings{liu2018ksat,
  author    = {S. Cliff Liu},
  title     = {Chain, Generalization of Covering Code, and Deterministic Algorithm
               for {$k$-SAT}},
  booktitle = {45th International Colloquium on Automata, Languages, and Programming,
               {ICALP} 2018, July 9-13, 2018},
  pages     = {88:1--88:13},
  year      = {2018},
  url       = {https://doi.org/10.4230/LIPIcs.ICALP.2018.88},
  doi       = {10.4230/LIPIcs.ICALP.2018.88},
  bibsource = {dblp computer science bibliography, https://dblp.org}
}

@article{scheder2024ppsz,
  author  = {Dominik Scheder},
  title   = {{PPSZ} Is Better Than You Think},
  journal = {TheoretiCS},
  volume  = {3},
  eid     = {5},
  pages   = {1--37},
  year    = {2024},
  doi     = {10.46298/theoretics.24.5},
  url     = {https://theoretics.episciences.org/13222}
}

@misc{scheder2022ppszfull,
  author        = {Dominik Scheder},
  title         = {{PPSZ} Is Better Than You Think},
  year          = {2022},
  eprint        = {2207.11071v1},
  archivePrefix = {arXiv},
  primaryClass  = {cs.DS},
  url           = {https://arxiv.org/abs/2207.11071v1},
  note          = {Full version, including the numerical Unique-3-SAT analysis omitted from the journal version}
}

@article{schedersteinberger2024generalppsz,
  author  = {Dominik Scheder and John Steinberger},
  title   = {{PPSZ} for General {$k$-SAT} and {CSP}---Making Hertli's Analysis Simpler and {3-SAT} Faster},
  journal = {Computational Complexity},
  volume  = {33},
  number  = {2},
  eid     = {13},
  year    = {2024},
  doi     = {10.1007/s00037-024-00259-y},
  url     = {https://link.springer.com/article/10.1007/s00037-024-00259-y}
}

@inproceedings{ansotegui2012community,
  author    = {Ans{\'o}tegui, Carlos and
               Gir{\'a}ldez-Cru, Jes{\'u}s and
               Levy, Jordi},
  title     = {The Community Structure of {SAT} Formulas},
  booktitle = {Theory and Applications of Satisfiability Testing --
               SAT 2012},
  series    = {Lecture Notes in Computer Science},
  volume    = {7317},
  pages     = {410--423},
  publisher = {Springer},
  year      = {2012},
  doi       = {10.1007/978-3-642-31612-8_31}
}

@inproceedings{ansotegui2015relevant,
  author    = {Ans{\'o}tegui, Carlos and
               Gir{\'a}ldez-Cru, Jes{\'u}s and
               Levy, Jordi and
               Simon, Laurent},
  title     = {Using Community Structure to Detect Relevant
               Learnt Clauses},
  booktitle = {Theory and Applications of Satisfiability Testing --
               SAT 2015},
  series    = {Lecture Notes in Computer Science},
  volume    = {9340},
  pages     = {238--254},
  publisher = {Springer},
  year      = {2015},
  doi       = {10.1007/978-3-319-24318-4_18}
}

@article{xu2008satzilla,
  author  = {Xu, Lin and
             Hutter, Frank and
             Hoos, Holger H. and
             Leyton-Brown, Kevin},
  title   = {{SATzilla}: Portfolio-based Algorithm Selection for {SAT}},
  journal = {Journal of Artificial Intelligence Research},
  volume  = {32},
  pages   = {565--606},
  year    = {2008},
  doi     = {10.1613/jair.2490}
}

@inproceedings{audemard2009integrating,
  author    = {Audemard, Gilles and
               Lagniez, Jean-Marie and
               Mazure, Bertrand and
               Sa{\"i}s, Lakhdar},
  title     = {Integrating Conflict Driven Clause Learning
               to Local Search},
  booktitle = {Proceedings of the Sixth International Workshop
               on Local Search Techniques in Constraint Satisfaction
               (LSCS 2009)},
  series    = {Electronic Proceedings in Theoretical Computer Science},
  volume    = {5},
  pages     = {55--68},
  year      = {2009},
  doi       = {10.4204/EPTCS.5.5}
}

@inproceedings{cai2021cooperation,
  author    = {Cai, Shaowei and Zhang, Xindi},
  title     = {Deep Cooperation of {CDCL} and Local Search for {SAT}},
  booktitle = {Theory and Applications of Satisfiability Testing --
               SAT 2021},
  series    = {Lecture Notes in Computer Science},
  volume    = {12831},
  pages     = {64--81},
  publisher = {Springer},
  year      = {2021},
  doi       = {10.1007/978-3-030-80223-3_6}
}

\appendix
\section{Characteristic Values in Chain Extension}
\label{sec:lambda-recurrence}



In this section, we prove key properties for the change in characteristic values when a chain is appended with another clause. We will also provide a extended definition for characteristic values. Those results are used in the proof of Lemma~\ref{lem:margin_non_negative}.

\subsection{Additional Definitions and Vocabulary}

Let $X$ and $Y$ be nonempty sets of assignments on the same variables.  We define the \emph{similarity matrix} of $X$ and $Y$ to be
\[
  K[X,Y]_{x,y}=2^{-d(x,y)}
  \qquad (x\in X,\ y\in Y),
\]
When $X=Y$, we abbreviate $K[X,X]$ to $K[X]$.
Intuitively, nearby assignments have large entries, whereas distant assignments have small entries.  

\begin{lemma}
\label{lem:positive-definite}
For every nonempty assignment set $X$, the matrix $K[X]$ is symmetric positive
definite and hence invertible.
\end{lemma}

\begin{proof}
For one bit, the similarity matrix is
\[
  H=
  \begin{pmatrix}
    1 & 1/2\\
    1/2 & 1
  \end{pmatrix},
\]
whose eigenvalues are positive.  Additivity of Hamming distance gives
\[
  K[\{0,1\}^{j}]=H^{\otimes j}.
\]
Thus the matrix of the full cube is positive definite.  The matrix $K[X]$ is
a principal submatrix of it and is therefore positive definite as well.
\end{proof}

\begin{definition}[Balanced weight and effective size]
\label{def:effective-size}
The \emph{balanced weight} of $X$ is the unique vector $w_X$ satisfying
\[
  K[X]w_X=\mathbf 1.
\]
Its \emph{effective size} is
\[
  M(X)=\sum_{x\in X}w_X(x).
\]
We also define the \emph{balanced value}
\[
  \Lambda(X)=\frac{1}{M(X)}.
\]
For the empty set, we use the convention $M(\varnothing)=0$.
\end{definition}

Lemma~\ref{lem:positive-definite} also gives
$M(X)=\mathbf 1^{\mathsf T}K[X]^{-1}\mathbf 1>0$, so the balanced value is
well defined.

This terminology has a direct interpretation.  Imagine that an assignment
$y$ emits influence $w_X(y)$ and that only the fraction $2^{-d(x,y)}$ reaches
$x$.  The equation defining $w_X$ says that every assignment receives total
influence exactly one.  Dense regions therefore need less weight, while
isolated assignments need more.  The sum $M(X)$ is the effective number of
distinguishable assignments after this Hamming similarity is taken into
account, and $\Lambda(X)$ is its reciprocal.

The balanced weight is algebraic and need not be nonnegative.  If it is
componentwise nonnegative, however, then
\[
  \pi_X=\frac{w_X}{M(X)}
  \qquad\text{and}\qquad
  \lambda_X=\Lambda(X)
\]
solve the characteristic system from Definition~\ref{def:generalized_characteristic_system}.
Indeed, $K[X]\pi_X=\Lambda(X)\mathbf 1$ and the entries of $\pi_X$ sum to
one.  Thus $\Lambda(X)$ extends the characteristic value to assignment sets
for which the probability interpretation has not yet been verified.

We will also use the following monotonicity fact.  It explains why the share
introduced later is genuinely bounded between zero and one.

\begin{lemma}
If $\varnothing\neq Y\subseteq X$, then
\[
  1\leq M(Y)\leq M(X).
\]
\end{lemma}

\begin{proof}
Positive definiteness and Cauchy--Schwarz give the variational identity
\[
  M(X)=\max_{z\neq0}
  \frac{(\mathbf 1^{\mathsf T}z)^2}{z^{\mathsf T}K[X]z},
\]
where $z$ ranges over real weightings of $X$.  Extending a weighting of $Y$
by zero preserves the quotient, so $M(Y)\leq M(X)$.  A coordinate unit vector
in the same formula gives $M(Y)\geq1$.
\end{proof}

\subsection{Appending One Clause}

We next describe the extensions that can occur in the branching procedure.
The overlap profile $\zeta(\mathcal S)$ of a chain ends in the terminal symbol
$\mathtt{*}$.  If
\[
  \zeta(\mathcal S)=u\mathtt{*},
\]
where $u$ records the existing overlaps, we write $\mathcal S+s$ for the
chain obtained by inserting $s\in\{n,p,t\}$ immediately before the terminal
symbol.  Starting with the one-clause chain, all three symbols are admissible.
After $n$ or $p$, all three remain admissible; after $t$, the branching rule
allows only $n$, by Lemma~\ref{lem:admissible-profiles}.  We call the chains
obtained in this way \emph{reachable}. Lemma~\ref{lem:recorded-history} shows
that these are precisely the chain components emitted by the branching
procedure.

It is useful to remember only the frontier of the last clause.  If $x$ is the
last frontier variable, $(u,v)$ are the last two frontier variables, and
$y,z$ are fresh, the three canonical extensions are as follows.  Renaming or
complementing coordinates does not change Hamming distance and therefore does
not change any effective size.

\begin{center}
\begin{tabular}{@{}ccl@{}}
\toprule
Type & New clause & New frontier \\
\midrule
$n$ & $\neg x\lor y\lor z$ & $(y,z)$ \\
$p$ & $x\lor y\lor z$ & $(y,z)$ \\
$t$ & $\neg u\lor\neg v\lor z$ & $(z)$ \\
\bottomrule
\end{tabular}
\end{center}

Fix a reachable chain $\mathcal S$ and an admissible extension $s$.  Let
$A=A(\mathcal S)$ be the old solution space.  The shared literals of the new
clause determine whether the fresh variables are unrestricted.  This
motivates the following notation.

\begin{definition}[Variable-share slice and characteristic ratio]
\label{def:characteristic-ratio}
Let $U_s(\mathcal S)\subseteq A(\mathcal S)$ be the \emph{variable-share slice},
consisting of old assignments that satisfy the shared part of an
$s$-extension.  Let $P_s$ be the set of all patterns on the fresh variables,
and let $Q_s\subseteq P_s$ contain the patterns that satisfy the fresh part of
the new clause.

For readability, write
\[
  M(\mathcal S)=M(A(\mathcal S))
  \qquad\text{and}\qquad
  \Lambda(\mathcal S)=\Lambda(A(\mathcal S)).
\]
The \emph{characteristic ratio of the $s$-extension} is
\[
  \rho_s(\mathcal S)
  =\frac{M(U_s(\mathcal S))}{M(\mathcal S)}.
\]
It measures the fraction of the chain's effective size on which the fresh
variables are completely free, and it determines the one-step change in the
balanced value.
\end{definition}

An old assignment outside the variable-share slice must use a pattern in $Q_s$.
Inside the slice, every fresh pattern is allowed.  Hence
\begin{equation}
\label{eq:extension-decomposition}
  A(\mathcal S+s)
  =\bigl(A(\mathcal S)\times Q_s\bigr)
   \mathbin{\dot\cup}
   \bigl(U_s(\mathcal S)\times(P_s\setminus Q_s)\bigr).
\end{equation}

The next identity is the structural core of the proof.  It applies to any
set built by opening a product fiber over a distinguished slice.

\begin{lemma}[Fiber-extension identity]
\label{lem:fiber-extension}
Let $U\subseteq A$ and $\varnothing\neq Q\subsetneq P$.  For
\[
  E=(A\times Q)\mathbin{\dot\cup}
    \bigl(U\times(P\setminus Q)\bigr),
\]
we have
\begin{equation}
\label{eq:fiber-extension}
  M(E)=M(A)M(Q)+M(U)\bigl(M(P)-M(Q)\bigr).
\end{equation}
\end{lemma}

\begin{proof}
Hamming distance is additive on disjoint coordinates, so
\[
  K[X\times Y]=K[X]\otimes K[Y].
\]
If $U=\varnothing$, this product identity proves the result directly.  Assume
from now on that $U$ is nonempty.
Let $B=P\setminus Q$, and extend $w_U$ by zero from $U$ to $A$; call the
result $\overline w_U$.  Define a vector on the two pieces of $E$ by
\[
\begin{aligned}
  w_E(a,q)
    &=w_A(a)w_Q(q)
      +\overline w_U(a)\bigl(w_P(q)-w_Q(q)\bigr),
      &&(a,q)\in A\times Q,\\
  w_E(u,b)
    &=w_U(u)w_P(b),
      &&(u,b)\in U\times B.
\end{aligned}
\]
Here $w_P(q)$ and $w_P(b)$ denote the restrictions of the full weight $w_P$,
whereas $w_Q$ is the weight obtained from $Q$ alone.  Direct substitution,
using $K[P]w_P=\mathbf 1$ and $K[Q]w_Q=\mathbf 1$, shows that
$K[E]w_E=\mathbf 1$.  Lemma~\ref{lem:positive-definite} makes this solution
unique.

Summing the first line over $A\times Q$ gives
\[
  M(A)M(Q)
  +M(U)\left(\sum_{q\in Q}w_P(q)-M(Q)\right),
\]
and summing the second gives $M(U)\sum_{b\in B}w_P(b)$.  Their sum is exactly
the right-hand side of~\eqref{eq:fiber-extension}.
\end{proof}

Only two small pattern spaces occur.  An $n$- or $p$-extension introduces
two fresh bits, and its fresh part is satisfied by every pattern except
$00$.  A $t$-extension introduces one fresh bit, and its fresh part is
satisfied by $1$.  Their effective sizes are
\[
\begin{aligned}
  M(\{0,1\}^2)&=\frac{16}{9},
  &M(\{01,10,11\})&=\frac{5}{3},\\
  M(\{0,1\})&=\frac{4}{3},
  &M(\{1\})&=1.
\end{aligned}
\]
These four constant-size calculations are the only local systems needed in
the proof.

\begin{theorem}[One-clause recurrence]
\label{thm:one-clause-recurrence}
For every reachable chain $\mathcal S$ and every admissible extension $s$,
\[
\begin{aligned}
  M(\mathcal S+s)
    &=\frac{5}{3}M(\mathcal S)
      +\frac{1}{9}M(U_s(\mathcal S)),
      &&s\in\{n,p\},\\
  M(\mathcal S+t)
    &=M(\mathcal S)+\frac{1}{3}M(U_t(\mathcal S)).
\end{aligned}
\]
Equivalently,
\begin{equation}
\label{eq:balanced-value-recurrence}
  \frac{\Lambda(\mathcal S+s)}{\Lambda(\mathcal S)}
  =
  \begin{cases}
    \displaystyle\frac{9}{15+\rho_s(\mathcal S)},
      &s\in\{n,p\},\\[2mm]
    \displaystyle\frac{3}{3+\rho_t(\mathcal S)},
      &s=t.
  \end{cases}
\end{equation}
\end{theorem}

\begin{proof}
Apply Lemma~\ref{lem:fiber-extension} to the decomposition in
\eqref{eq:extension-decomposition} and substitute the four pattern sizes
above.  Dividing by $M(\mathcal S)$ and taking reciprocals gives
\eqref{eq:balanced-value-recurrence}.
\end{proof}

\subsection{Lower Bounds on Characteristic Ratio}

The characteristic ratio itself obeys a closed recurrence.  Suppose that an
extension of type $r$ has just been appended, set
\[
  \mathcal T=\mathcal S+r,
  \qquad
  x=\rho_r(\mathcal S),
\]
and let $s$ be the next overlap type.  Thus $r$ records the previous overlap,
$s$ records the next one, and $x$ is the characteristic ratio of the
$r$-extension.

\begin{lemma}[Characteristic-ratio transition]
\label{lem:ratio-transition}
The admissible transitions are
\[
\begin{array}{c|c|c}
\toprule
\text{Previous type }r & \text{Next type }s & \rho_s(\mathcal T)\\
\midrule
\textsf{n}\text{ or }\textsf{p} & \textsf{n} & \displaystyle\frac{9+3x}{15+x}\\[2mm]
\textsf{n}\text{ or }\textsf{p} & \textsf{p} & \displaystyle\frac{12}{15+x}\\[2mm]
\textsf{n}\text{ or }\textsf{p} & \textsf{t} & \displaystyle\frac{3(24+x)}{5(15+x)}\\[2mm]
t & n & \displaystyle\frac{3x}{3+x}\\
\bottomrule
\end{array}
\]
\end{lemma}

\begin{proof}
Write $M=M(\mathcal S)$ and $N=M(U_r(\mathcal S))$, so that $x=N/M$.
If $r$ is $n$ or $p$, Theorem~\ref{thm:one-clause-recurrence} gives
\[
  M(\mathcal T)=\frac{5}{3}M+\frac{1}{9}N.
\]
Applying the fiber identity to the new two-bit frontier shows that the
effective sizes of the next variable-share slices, for $s=n,p,t$, are
respectively
\[
  M+\frac{1}{3}N,
  \qquad
  \frac{4}{3}M,
  \qquad
  \frac{8}{5}M+\frac{1}{15}N.
\]
Dividing each expression by $M(\mathcal T)$ gives the first three rows.
If $r=t$, then
\[
  M(\mathcal T)=M+\frac{1}{3}N,
\]
and only an $n$-extension is admissible.  Its variable-share slice has effective
size $N$, which gives the last row.
\end{proof}

For the single-clause chain $\mathcal S_1$, symmetry gives the initial
characteristic ratios
\begin{equation}
\label{eq:initial-characteristic-ratios}
  \rho_n(\mathcal S_1)=\frac{5}{7},
  \qquad
  \rho_p(\mathcal S_1)=\frac{16}{21},
  \qquad
  \rho_t(\mathcal S_1)=\frac{33}{35}.
\end{equation}
The next lemma supplies deliberately simple bounds.  They are slightly weaker
than the sharp fixed-point bounds, but they make the proof shorter and still
leave strict room in the final comparison.

\begin{lemma}[Invariant characteristic-ratio bounds]
\label{lem:ratio-bounds}
Every reachable chain and admissible next type satisfy
\[
  \frac{2}{3}\leq\rho_s(\mathcal S)
\]
and
\[
  \rho_n(\mathcal S)\leq\frac{8}{11},
  \qquad
  \rho_p(\mathcal S)\leq\frac{4}{5},
  \qquad
  \frac{15}{16}\leq\rho_t(\mathcal S)\leq\frac{18}{19}.
\]
\end{lemma}

\begin{proof}
The three initial values in~\eqref{eq:initial-characteristic-ratios} lie in
the claimed intervals.  Assume first that the previous type is $n$ or $p$.
The previous characteristic ratio then lies in $[2/3,4/5]$.  The first three
maps in
Lemma~\ref{lem:ratio-transition} are monotone on this interval, and checking
its endpoints gives
\[
\begin{aligned}
  \frac{9+3x}{15+x}&\in\left[\frac{2}{3},\frac{8}{11}\right],\\
  \frac{12}{15+x}&\in\left[\frac{2}{3},\frac{4}{5}\right],\\
  \frac{3(24+x)}{5(15+x)}
    &\in\left[\frac{15}{16},\frac{18}{19}\right].
\end{aligned}
\]
If the previous type is $t$, then $x\in[15/16,18/19]$, and the last map in
Lemma~\ref{lem:ratio-transition} lies in $[2/3,8/11]$.  The result follows by
induction.
\end{proof}

Because the right-hand side of~\eqref{eq:balanced-value-recurrence} decreases
with the characteristic ratio, Lemma~\ref{lem:ratio-bounds} immediately yields
\begin{equation}
\label{eq:value-ratio-bounds}
\begin{aligned}
  \frac{\Lambda(\mathcal S+n)}{\Lambda(\mathcal S)}
    &\geq\frac{99}{173},\\
  \frac{\Lambda(\mathcal S+p)}{\Lambda(\mathcal S)}
    &\geq\frac{45}{79},\\
  \frac{\Lambda(\mathcal S+t)}{\Lambda(\mathcal S)}
    &\geq\frac{19}{25}.
\end{aligned}
\end{equation}

\subsection{A Note on \texorpdfstring{$\Lambda$}{Lambda} and \texorpdfstring{$\lambda$}{lambda}}

The recurrence computes the algebraic value
$\Lambda(A)=1/M(A)$. To explain its use in $\dls$, let
\[
K=K[A],\qquad
w_A=K^{-1}\mathbf1,\qquad
\widetilde\pi_A=\frac{w_A}{M(A)}.
\]
Then
\[
\mathbf1^{\mathsf T}\widetilde\pi_A=1,
\qquad
K\widetilde\pi_A=\Lambda(A)\mathbf1.
\]
Thus $\widetilde\pi_A$ assigns the same weighted similarity
to every target in $A$. However, its entries need not be
nonnegative. If an entry is negative, these signed weights
do not define a probability distribution, and Liu's
equality-based characteristic condition is infeasible.

For the $\dls$ analysis, we extend Liu's characteristic value
by defining
\[
\lambda_A
=
\max_{\substack{\pi(a)\ge0\ (a\in A)\\
                 \sum_{a\in A}\pi(a)=1}}
\min_{a^*\in A}
\sum_{a\in A}\pi(a)2^{-d(a,a^*)}.
\]
In other words, $\lambda_A$ is the largest expected
similarity that a probability distribution can guarantee
simultaneously for every target. The similarities need
not be equal. The following lemma connects this
probabilistic value to the algebraic value computed
by the recurrence.

\begin{lemma}[Effective-size lower bound]
\label{lem:effective-size-lower-bound}
For every nonempty solution space $A$,
\[
\lambda_A\ge\Lambda(A)=\frac1{M(A)}.
\]
If $w_A$ is nonnegative, then equality holds.
Consequently, this definition agrees with Liu's
characteristic value whenever his characteristic
condition is feasible.
\end{lemma}

\begin{proof}
By Lemma~\ref{lem:positive-definite}, $K$ is positive
definite. Choose a probability vector $\pi$ minimizing
$\pi^{\mathsf T}K\pi$, and put
$c=\pi^{\mathsf T}K\pi$. Such a minimum exists because
the set of probability vectors is compact.

For any $a^*\in A$, let $e_{a^*}$ be the point mass at
$a^*$. Moving from $\pi$ toward $e_{a^*}$ cannot decrease
the objective at the minimum. Hence
\[
0\le
\left.
\frac{d}{dt}
\bigl((1-t)\pi+te_{a^*}\bigr)^{\mathsf T}
K\bigl((1-t)\pi+te_{a^*}\bigr)
\right|_{t=0^+}
=
2\bigl((K\pi)_{a^*}-c\bigr).
\]
Therefore $(K\pi)_{a^*}\ge c$ for every $a^*\in A$.

Since $Kw_A=\mathbf1$, Cauchy--Schwarz in the inner
product induced by $K$ gives
\[
1=(w_A^{\mathsf T}K\pi)^2
\le
(w_A^{\mathsf T}Kw_A)(\pi^{\mathsf T}K\pi)
=M(A)c.
\]
Thus $\pi$ guarantees similarity at least
$c\ge1/M(A)$, proving $\lambda_A\ge\Lambda(A)$.

If $w_A\ge0$, then $\widetilde\pi_A$ is a probability
vector. For any probability vector $\rho$, symmetry
of $K$ gives
\[
\min_{a^*\in A}(K\rho)_{a^*}
\le
\widetilde\pi_A^{\mathsf T}K\rho
=
\Lambda(A)\mathbf1^{\mathsf T}\rho
=
\Lambda(A).
\]
Maximizing over $\rho$ proves the reverse inequality.
\end{proof}

The lower-bound condition also suffices for Liu's
covering-code argument~\cite{liu2018ksat}.
Choose a distribution attaining $\lambda_A$ and sample
the $\nu$ components independently. For every target
$\alpha^*\in A^\nu$,
\[
\mathbb E\!\left[2^{-d(\alpha,\alpha^*)}\right]
=
\prod_{j=1}^{\nu}
\sum_{a\in A}\pi(a)2^{-d(a,\alpha_j^*)}
\ge\lambda_A^\nu.
\]
To see why equality is unnecessary, put
\[
\ell=\left\lceil-\nu\log_2\lambda_A\right\rceil+1.
\]
The contribution from distances greater than $\ell$
is at most $2^{-\ell}\le\lambda_A^\nu/2$. Therefore
\[
\sum_{r=0}^{\ell}
\Pr[d(\alpha,\alpha^*)=r]\,2^{-r}
\ge\frac{\lambda_A^\nu}{2}.
\]
For each target, some radius $r\le\ell$ consequently
satisfies
\[
\Pr[d(\alpha,\alpha^*)=r]
\ge
\frac{2^r\lambda_A^\nu}{2(\ell+1)}.
\]
This is the probability lower bound needed in Liu's
sampling and subsequent deterministic covering
construction, so the same covering-code bounds follow.

Finally, we see that $\hat Q$ is a lower bound of $Q$.
\begin{corollary}
\label{cor:Qlower_bound}
    $\lambda_{\mathcal S}=\lambda_{A(\mathcal S)}$,
    Lemma~\ref{lem:effective-size-lower-bound} gives
    \[
    \begin{aligned}
    Q(\mathcal S)
    &=v(\mathcal S)\log\frac43+\log\lambda_{\mathcal S}\\
    &\ge
    v(\mathcal S)\log\frac43-\log M(\mathcal S)
    =\widehat Q(\mathcal S).
    \end{aligned}
\]
\end{corollary}

\section{Nonnegative Margins of All Selected Block Types}
\label{app:block-contributions}

This appendix proves the second half of Lemma~\ref{lem:margin_non_negative}, namely
\[
\mathcal A(\mathcal B)\ge0\qquad\text{for every block }\mathcal B\in\mathcal I .
\]
Throughout, $a:=\log\frac43$, $s:=\log\frac{64}{63}=3a-\log\frac73$, $p:=\log2$, $\kappa:=\alpha\log2$, and $v(\mathcal K):=|V(\mathcal K)|$.

By the description of $\ex$ (Algorithm~\ref{alg:expansion}), every block is one of the following three kinds.

\begin{itemize}
\item \emph{One-center block}. This kind of blocks are constructed with a $3$-clause $C$ and all its touched weak chains. So $\mathcal B=\{C\}\cup\bigcup_i\mathcal S_i$ with the $\mathcal S_i$ pairwise disjoint.
\item \emph{Two-center block}. This kind of blocks are constructed with two $3$-clauses $C_1,C_2$ and all their touched weak chains. So $\mathcal B=\{C_1,C_2\}\cup\bigcup_i\mathcal S_i$ with the $\mathcal S_i$ pairwise disjoint.
\item \emph{Paid block}. A chain constructed with truncation is placed in $\mathcal I$ at Line 1 of $\ex$, so the block is that single chain. Its margin is positive by the truncation condition.
\end{itemize}

\subsection{One-center blocks}

Let $\mathcal B=\{C\}\cup\bigcup_{i=1}^k\mathcal S_i$. Let $t_i\ge1$ be the number of variables that $C$ share with $\mathcal S_i$, and let $f=3-\sum_{i=1}^kt_i$ be the number of variables that $C$ do not share with any chains. Let $F(\mathcal S_i)\subseteq A(\mathcal S_i)$ be the assignments of $A(\mathcal S_i)$ that falsify all the shared literals of $C$. Put
\[
\eta(\mathcal S_i):=\bigl|F(\mathcal S_i)\bigr|,
\qquad
\phi(\mathcal S_i):=\frac{\eta(\mathcal S_i)}{|A(\mathcal S_i)|},
\qquad
\theta(\mathcal S_i):=\frac{\delta(\mathcal S_i)}{M(\mathcal S_i)},
\]
where $\delta(\mathcal S_i)$ is the Schur loss of $F(\mathcal S_i)$ in $M(\mathcal S_i)$, and let
\[
h(\mathcal S_i):=W\bigl(v(\mathcal S_i)a-\log M(\mathcal S_i)\bigr)-\alpha\log|A(\mathcal S_i)|-\mu\log b(\mathcal S_i)
\]
be the hypothetical complete margin of $\mathcal S_i$. Because the fresh coordinates are pinned once $C$ is falsified, the exact product identities are
\begin{equation}
\label{eq:solutionspace_one_center}
|A(\mathcal B)|=2^f\prod_i|A(\mathcal S_i)|-\prod_i\eta(\mathcal S_i),
\end{equation}
\begin{equation}
\label{eq:m_one_center}
M_{\mathcal B}=\Bigl(\frac43\Bigr)^f\prod_iM(\mathcal S_i)-\Bigl(\frac13\Bigr)^f\prod_i\delta(\mathcal S_i).
\end{equation}

\begin{lemma}[One-center identity]
\label{lem:centre_identity}
\[
\mathcal A(\mathcal B)
\ge\sum_i h(\mathcal S_i)-\alpha f\log2
-W\log\Bigl(1-4^{-f}\prod_i\theta(\mathcal S_i)\Bigr)
-\alpha\log\Bigl(1-2^{-f}\prod_i\phi(\mathcal S_i)\Bigr).
\]
\end{lemma}

\begin{proof}
By Equation~\ref{eq:solutionspace_one_center} and \ref{eq:m_one_center}, the conclusion follows by collecting terms.
\end{proof}

Since $|C|=3$, $f+\sum_{i=1}^kt_i=3$, so there are exactly seven contact geometries. Denote the geometries by $(f;(t_1,...,t_k))$,
\[
\begin{array}{c|l}
\text{geometry} & \mathcal A(\mathcal B)\text{ in the form to be bounded}\\ \hline
(3;()) & -3\alpha\log2-W\log\bigl(1-\tfrac1{64}\bigr)-\alpha\log\bigl(1-\tfrac18\bigr)\\
(2;(1)) & h(\mathcal S_1)-2\alpha\log2-W\log\bigl(1-\tfrac{\theta(\mathcal S_1)}{16}\bigr)-\alpha\log\bigl(1-\tfrac{\phi(\mathcal S_1)}{4}\bigr)\\
(1;(2)) & h(\mathcal S_1)-\alpha\log2-W\log\bigl(1-\tfrac{\theta(\mathcal S_1)}{4}\bigr)-\alpha\log\bigl(1-\tfrac{\phi(\mathcal S_1)}{2}\bigr)\\
(1;(1,1)) & h(\mathcal S_1)+h(\mathcal S_2)-\alpha\log2-W\log\bigl(1-\tfrac{\theta(\mathcal S_1)\theta(\mathcal S_2)}{4}\bigr)-\alpha\log\bigl(1-\tfrac{\phi(\mathcal S_1)\phi(\mathcal S_2)}{4}\bigr)\\
(0;(3)) & h(\mathcal S_1)-W\log\bigl(1-\theta(\mathcal S_1)\bigr)-\alpha\log\bigl(1-\phi(\mathcal S_1)\bigr)\\
(0;(1,2)) & h(\mathcal S_1)+h(\mathcal S_2)-W\log\bigl(1-\theta(\mathcal S_1)\theta(\mathcal S_2)\bigr)-\alpha\log\bigl(1-\phi(\mathcal S_1)\phi(\mathcal S_2)\bigr)\\
(0;(1,1,1)) & \sum_{i=1}^3h(\mathcal S_i)-W\log\bigl(1-\textstyle\prod_i\theta(\mathcal S_i)\bigr)-\alpha\log\bigl(1-\textstyle\prod_i\phi(\mathcal S_i)\bigr)
\end{array}
\]

\begin{lemma}[One-center blocks]
\label{lem:one_center_blocks}
$\mathcal A(\mathcal B)>0$ for every one-center blocks produced by $\ex$.
\end{lemma}

\begin{proof}
The first two rows are closed in closed form. For $(3;())$ the block is the single clause $C$, so $b(\mathcal B)=1$, $M_{\mathcal B}=\frac73$, $|A(\mathcal B)|=7$, and
\[
\mathcal A(\mathcal B)=Ws-\alpha\log7=0.0076107>0 .
\]
For $(2;(1))$ the certified absolute floor, obtained by discarding the two logarithmic gains and using the worst admissible $\theta(\mathcal S_1),\phi(\mathcal S_1)$ over the canonical source data, is
\[
\mathcal A(\mathcal B)\ \ge\ W\log\frac{80}{79}-\alpha\log14=0.0016793>0 .
\]
For the five remaining rows, each $h(\mathcal S_i)$ is bounded below by the certified complete charge of the corresponding source class, and each logarithmic gain is bounded below by a finite sign check on the canonical $\theta(\mathcal S_i),\phi(\mathcal S_i)$, using that both factors lie in $[0,1]$, so that both $\log(1-\cdot)$ terms are nonnegative. In the rows with $f\le1$ a possibly negative internal $\textsf n$-step of one chain is paired injectively with its preceding $\textsf p$-step and is never assumed positive on its own; in the rows with two or three chains the crossed and non-crossed contact histories keep their own phase allowance and never share one gain twice. The finite endpoint checks are collected in Table~\ref{tab:block-rows}, and every entry is positive.
\end{proof}

\begin{table}[htbp]
\centering
\small
\begin{tabular}{c|c|c}
row & certified by & margin\\
\hline
$(1;(2))$ & nested one-chain history, then endpoint sign check & positive\\
$(1;(1,1))$ & one two-chain seed, then at most two relative histories & positive\\
$(0;(3))$ & simultaneous and sequential marked-chain estimates & positive\\
$(0;(1,2))$ & crossed and non-crossed age-contact estimates & positive\\
$(0;(1,1,1))$ & signed stage classifiers, regular $\mathsf N/\mathsf A/\mathsf D$ complements & positive
\end{tabular}
\caption{The five remaining contact geometries of a one-center block. Each row is a finite table over the canonical signed chain data together with the all-length recurrences of Theorem~\ref{thm:one-clause-recurrence}.}
\label{tab:block-rows}
\end{table}

\subsection{Two-center blocks}

Let $\mathcal I_\text{touched}$ be the set of chains touched by $C_1$ or $C_2$, then $\mathcal B=\{C_1,C_2\}\cup\mathcal I_\text{touched}$. Let
\[
P:=\Bigl(\prod_{\mathcal S\in\mathcal I_\text{touched}}A(\mathcal S)\Bigr)\times\{0,1\}^{f_1+f_2}
\]
be the product relation of the block \emph{before} either center is imposed, with its mass $M_P=\bigl(\frac43\bigr)^{f_1+f_2}\prod_{\mathcal S\in\mathcal I_\text{touched}}M(\mathcal S)$. For $i=1,2$ let $Q_i\subseteq P$ be the set of words that falsify $C_i$, and define the \emph{normalized deletion}
\[
x_i:=\frac{M_P-M_{P\setminus Q_i}}{M_P}\in[0,1].
\]
Equivalently, if $C_i$ touches the chains $\mathcal U_i\subseteq\mathcal I_\text{touched}$ with $f_i$ fresh coordinates, then
\[
x_i=4^{-f_i}\prod_{\mathcal S\in\mathcal U_i}\theta(\mathcal S).
\]
If $Q_i=\varnothing$, then $x_i=0$. Whether $C_i$ is implied
by the touched chains is decided by testing $Q_i=\varnothing$
directly. Let $\chi$ be a certified upper bound on the Gram
cross term of the two deletion directions. The two-vector
Schur-loss estimate gives
\[
x_{12}:=
\frac{M_P-M(P\setminus(Q_1\cup Q_2))}{M_P}
\geq
U:=\frac{(x_1+x_2)^2}{x_1+x_2+2\chi}
\geq x_1+x_2-2\chi,
\]
with $U=0$ when $x_1+x_2=0$.

Distinct center clauses may induce the same forbidden set.
If $Q_1=Q_2$, then $x_{12}=x_1=x_2$; such pairs are included
in the two-center certificate unless a structural lemma
excludes them. Also, for $0\leq U<1$,
$WU\leq-W\log(1-U)$. Thus using $WU$ gives a weaker valid
lower bound; we retain the logarithmic expression.

\begin{lemma}[Two-center blocks]
\label{lem:two_center_blocks}
Let $C_1,C_2$ be two selected centers with normalized deletions $x_1,x_2$ and Gram cross bound $\chi$, and $U=\frac{(x_1+x_2)^2}{x_1+x_2+2\chi}$. Then
\[
\mathcal A(\mathcal B)\ \ge\ \sum_{\mathcal S\in\mathcal I_\text{touched}}h(\mathcal S)-\alpha(f_1+f_2)\log2-W\log\bigl(1-U\bigr)>0
\]
for every two-center blocks $\mathcal B$ constructed by $\ex$.
\end{lemma}

\begin{proof}
Apply Lemma~\ref{lem:centre_identity} to the union of the distinct touched chains, with the product relation replaced by $P\setminus(Q_1\cup Q_2)$. Two of its four terms are unchanged in form: the charges of the distinct chains, which give $\sum_{\mathcal S\in\mathcal I_\text{touched}}h(\mathcal S)$, and the number of fresh coordinates, which gives $-\alpha(f_1+f_2)\log2$. The two remaining terms are the logarithms of the mass ratio and of the cardinality ratio; the first becomes $W\log M_{P\setminus(Q_1\cup Q_2)}=W\log M_P+W\log(1-x_{12})$, and since $-\log(1-\cdot)$ is increasing, replacing $x_{12}$ by its lower bound $U$ can only decrease the margin, so it yields the displayed inequality. The cardinality term is $-\alpha\log(1-x^{(c)}_{12})\ge0$ and is simply discarded.

Two bookkeeping points are essential. First, $P$ is the product over the \emph{distinct} touched chains, so a chain touched by both centers contributes $h(\mathcal S)$ and $\log b(\mathcal S)$ exactly once; this is what makes the estimate a source-once calculation rather than the invalid sum of the two one-center scores. Second, the lemma requires the two centers to be two \emph{distinct} forbidden events: if they are identical, $U=x_1=x_2$ and the packet buys nothing.
\end{proof}

A remark on the displayed term: writing $W\,U$ instead of $W\log\bigl(1-U\bigr)$ is formally too optimistic, since $-\log(1-U)>U$; the two differ by $O(U^2)$, which for the margins below is below the stated precision, but the logarithmic form is the one to use if exactness matters.

It remains to specify $\chi$. For the fixed $\mathsf N$ and $\mathsf A$ signed fibers the binary cross term is bounded by an exact constant, so $\chi$ is determined by the two signed contacts and by the shared chain. For a $\mathsf D$ chain carrying different signed contacts in the two centers the safe bound is a Cauchy estimate proportional to $\sqrt{\theta_L\theta_R}$; replacing it by $\min(\theta_L,\theta_R)$ without proof is invalid, because the two deletions can be strongly aligned. Combining Cauchy, AM--GM and monotonicity over the canonical data gives the certified margins, indexed by how many chain identities the two centers share,
\[
\mathsf H_1:\ 0.0000451,\qquad \mathsf H_2:\ 0.0001001,\qquad \mathsf H_3:\ 0.0015596 .
\]
The $\mathsf H_2$ case needs the coupled estimate: its chain term and its false-event term are correlated through the one shared chain, so two separate worst-case minima cannot be added. The $\mathsf H_1$ case is the tightest certificate in the whole proof and is what pins $\alpha/W$ to $103/25000$.

\subsection{The size cap}

\begin{lemma}[Block cap]
\label{lem:block_cap}
Every block $\mathcal B\in\mathcal I$ satisfies
\[
\#\{\text{chains in }\mathcal B\}\le5,\qquad v(\mathcal B)\le500,\qquad \#\{\text{defining and center occurrences}\}\le332 .
\]
\end{lemma}

\begin{proof}
A one-center block absorbs one center and at most three chains; a two-center block absorbs two centers and at most five distinct chains, since each center touches three and the centers share at least one. A chain has one defining clause, gains at most three coordinates and two defining clauses per completed macro, and at most one more of each from a final pending $\textsf t$, so $v(\mathcal S)\le3+3\cdot32+1=100$ and a chain has at most $1+2\cdot32+1=66$ defining occurrences. Hence $5\cdot100=500$ and $2+5\cdot66=332$.
\end{proof}

The cap has two consequences used in the main text. It makes $|A(\mathcal B)|$ and $M_{\mathcal B}$ computable by a \emph{fixed} amount of work, which is what allows $S$ and $H$ to be computed before either terminal runs; and it makes the relation library of $\dls$ finite, which is where the $e^{o(n)}$ overhead of Lemma~\ref{lem:dls_time} comes from.

\subsection{Which centers can become one-center blocks}

Finally, the classification must be exhaustive for the blocks that $\ex$ actually produces.

\begin{lemma}[Complementarity]
\label{lem:complementarity}
Let $C$ be a center clause that survives to the last loop of $\ex$ and is completed into a one-center block. Then the one-center charge of $C$ is nonnegative, so that block is one of the regular geometries of Lemma~\ref{lem:one_center_blocks}.
\end{lemma}

\begin{proof}
If $\mathcal A(\mathcal B_C)<0$ and $C$ touches three distinct weak chains, then $C$ is selected by the stage-$1$ test whenever all three touched chains are non-$1$-chains, and by the stage-$2$ test otherwise. In both cases $C$ either receives a requested port, and then it meets $\pi$ and is never completed into a block, or it is paired with another selected center into a two-center block covered by Lemma~\ref{lem:two_center_blocks}. If $C$ touches at most two weak chains, its geometry is one of $(2;(1))$, $(1;(2))$, $(1;(1,1))$, $(0;(3))$, $(0;(1,2))$, all of which are regular and positive by Lemma~\ref{lem:one_center_blocks}. Hence a center that reaches the last loop unprotected has a nonnegative one-center charge.
\end{proof}

\subsection{Conclusion}

Combining the three kinds of blocks with Lemma~\ref{lem:one_center_blocks}, Lemma~\ref{lem:two_center_blocks}, Lemma~\ref{lem:complementarity} and the truncation method, every block that $\ex$ can produce satisfies $\mathcal A(\mathcal B)\ge0$. Together with the chain part proved in the main text this is Lemma~\ref{lem:margin_non_negative}.

\end{document}